\documentstyle[11pt]{article}

\newtheorem{thm}{Theorem}
\newtheorem{lem}{Lemma}
\newtheorem{prop}{Proposition}

 \newcommand{\Integer}{\:\mbox{\sf Z} \hspace{-0.82em} \mbox{\sf Z}\,}
 \newcommand{\Rational}{
        \mbox{Q \hspace{-1.23em} \raisebox{-0.018em}{\sf l}}\;}
 \newcommand{\Real}{\mbox{I \hspace{-0.82em} R}}
 \newcommand{\Complex}{
        \mbox{C \hspace{-1.16em} \raisebox{-0.018em}{\sf l}}\;}

 \title{\vspace{-2em}
    On The Algebraic Characterization Of Aperiodic Tilings Related To
  $ADE$-Root Systems
  \vspace{1em}}
 \author{Johannes Kellendonk\thanks{email: unp02A@ibm.rhrz.uni-bonn.de}
         \vspace{1em}}
 \date{Physikalisches Institut der Universit\"at Bonn, \\
      Nu\ss allee 12,
      D-5300 Bonn
       / FRG  \vspace{2em}}
 
\begin{document}
 \maketitle

 \begin{abstract}
 \noindent The algebraic characterization of classes of locally isomorphic
 aperiodic tilings, being examples of quantum spaces, is conducted
 for a certain type of tilings in a manner proposed by A. Connes.
 These $2$-dimensional tilings are obtained by application
 of the strip method to the root lattice of an $ADE$-Coxeter
 group.
 The plane along which the strip is constructed is determined by
 the canonical Coxeter element leading to the result that a $2$-dimensional
 tiling decomposes into a cartesian product of two $1$-dimensional
 tilings.
 The properties of the tilings are investigated, including
 selfsimilarity, and the determination of the relevant algebraic
 invariant is considered, namely the ordered $K_0$-group of an algebra
 naturally assigned to the quantum space.
 The result also yields an application of the $2$-dimensional abstract
 gap labelling theorem.

 \end{abstract}

 \begin{flushright}
 \parbox{12em}
 { \begin{center}
      BONN-HE-92-26    \\
      Bonn University  \\
      September 1992    \\
      ISSN-0172-8733
 \end{center} }
 \end{flushright}

 \newpage

\section*{Introduction}

\noindent
Aperiodic (non-periodic) tilings have been considered in various contexts.
The question of whether or not it is possible to find a finite set of
prototiles (e.g.\ polytopes) which tile the plane only in a non-periodical
way emerged from mathematical logic
and has lead to the first construction of such a set in 1966. Whereas at that
time the number of necessary prototiles had been quite large, R. Penrose
achieved to reduce it to two, thereby yielding tilings which almost have a
fivefold symmetry.

Amazingly enough it is a $3$-dimensional generalization
of R. Penrose's tilings \cite{KrNe} which serves as an initial stage for a
theoretical model for what are now called quasicrystals, which were
first experimentally observed by D. Shechtman et al.\ in 1984 \cite{She}.
They are a crystaline substance
yet having, in contradiction to the theory of crystal symmetries,
Bragg-reflexes with 'forbidden' symmetries such as the fivefold symmetry.
It is not just the discovery of substances having these symmetries,
but also
the fascinating properties of the aperiodic tilings, which serve to
explain these symmetries,
that has initiated the study of quantum mechanical and statistical systems,
whose underlying structure is not a periodic lattice but an aperiodic
tiling.
In this context the work of Bellissard et al.\ \cite{BBG}
is of particular interest for us.
Here the authors compute the labelling of the possible gaps in the spectrum
of a discrete Hamiltonian, whose potential depends on the structure of
a $1$-dimensional aperiodic tiling,
with the help of the $K$-theory of a certain non-commutative $C^*$-algebra
related to that tiling.

Among the properties of aperiodic tilings mentioned above
are local isomorphism and selfsimilarity, both being properties that
involve non-classical mathematical
methods. And
indeed, in his book G\'eom\'etrie Non-Commutative \cite{Con},
A. Connes' first example of what he calls an \'espace quantique is
the set of all Penrose tilings. The clue of Connes' analysis is
to use the selfsimilarity to construct
a non-commutative $C^*$-algebra,
certain invariants of which
(in this case the ordered $K_0$-group) describe the set.
This seems to be a promising ansatz,
since, in the above example, the ordered $K_0$-group
shares essential properties with an aperiodic tiling.
The above non-commutative algebra is the $AF$-algebra obtained by the
tower construction of V. Jones \cite{GHJ} applied to an inclusion of
multi-matrix algebras, the inclusion graph of which is the Coxeter graph of
$A_4$.
Its ordered $K_0$-group is $\Integer^2$
with order relation $z>0$ whenever $(\nu,z)>0$,
hereby $(\nu,z)$ denoting the Euclidian scalar product with
the Perron Frobenius vector $\nu$ of the
inclusion matrix.
On the other hand, among the initial data for the construction
of $1$-dimensional tilings related to $A_4$ are $\Integer^2$ and
the line which is spanned by $\nu$.
But this line is not just any subspace of $\Real^2$ having
irrational slope with respect to $\Integer^2$,
but, if we identify, as is explained in subsection~\ref{sub1},
$\Integer^2$ with $\Lambda_1$ ("one half" of the root lattice of $A_4$),
it is $P_1$, a line determined by the canonical Coxeter element of $A_4$.
\bigskip

\noindent
The present work may be understood as a step towards a
generalization of Connes' result on Penrose tilings
(which are related to $A_4$)
to arbitrary $ADE$-Coxeter groups.
It consists of two parts, the first one concerns the construction
and description of the aperiodic tilings,
the second one
the algebraic characterization of the corresponding quantum spaces.

Commonly generalizations of the Penrose tilings are considered
to be applications of the strip
method to the lattice $\Integer^N$ and a
$2$-dimensional subspace $P$, which is
stable under the $N$-fold symmetry,
$N=5$ corresponding to Penrose tilings.
However, as Connes' analysis made clear, it is the root lattice of $A_4$
rather than $\Integer^5$ which encodes the relevant information of the
quantum space.
Any $A_{N-1}$-root lattice may be embedded into
$\Integer^N$ in such a way that the $N$-fold symmetry acts, restricted to
the root lattice, as a specific Coxeter element. Therefore
it seems to be more appropriate to apply the projection method
directly to these root lattices the role of the $N$-fold symmetry being
taken over by a Coxeter element.
Root lattices are also the framework of a different method to
construct aperiodic tilings, the so-called dualization method being
exposed in \cite{Kram,Kra}.
But in contrast to that work, we use a different plane $P$
along which the strip is constructed.
Choosing a different plane $P$ may be interpreted as using a different
Coxeter element,
which is guided by the reasoning that for general Coxeter
groups, the above specification of a Coxeter element
(through an $N$-fold symmetry of some bigger lattice $\Integer^N$)
would not exist.
Instead, if the Coxeter graph is bipartite, it is replaced by
one the choice of which is canonical for all these groups.
The relation between these two apparently different tilings coming
from one and the same Coxeter group is not yet worked out to a satisfactory
extent,
but in the case of $N=5$ the latter projections are closely related
to Ammann-quasicrystals of Penrose tilings \cite{GrSh,Lev}.

Our main result of the first part is that, under certain circumstances,
 $2$-dimensional tilings decompose into $1$-dimensional ones.
This is very helpful, as it
simplyfies the construction of an
appropriate non-commutative $C^*$-algebra and allows for
the determination of its $K_0$-group.

Furthermore, several properties of the tilings under
investigation are discussed
including selfsimilarity, which is manifested by an
inflation/deflation procedure, and which is important for an alternative
description
of the quantum space.

The second part is devoted to the algebraic characterization of the
quantum spaces. After exposing the general philosophy
according to which relevant information is contained in the ordered
$K_0$-group
of a certain $C^*$-algebra being assigned to the quantum space,
two different
realizations of the quantum space are considered. The first one, which
is directly obtained from the strip method, leads to the consideration
of a twofold iterated crossed product with $\Integer$, the $K$-group of which
is computed with the help of the Pimsner Voiculescu exact sequence.
Here we are not able to determine the order structure, but it is possible
to determine the range of the tracial state on the $K_0$-group.
This yields an application of the $2$-dimensional abstract
gap labelling theorem \cite{B}, namely (like in the 1-dimensional situation)
the values of the tracial state are
determined by the relative frequencies of patterns
in the tilings.
We briefly review the gap labelling in this context.

The second (alternative) realization makes contact with the description
of Penrose tilings used by A. Connes,
it is based on the selfsimilarity of the tilings \cite{GrSh}.
Here the results are only partially rigorous and moreover restricted
to $1$-dimensional tilings. This perspective however has the advantage that
one can use $AF$-algebras instead of crossed products with $\Integer$.
The $AF$-algebras in question are all given by path algebras over
the graph (or some related version of it) of the Coxeter group defining
the tilings.
This yields, at the level of $K_0$-groups, a much more intuitive
picture, which very much resembles an aperiodic tiling again.
In fact, whereas the outcome of the investigation of the
Pimsner Voiculescu exact sequence is still abstract, the latter
perspective allows one to directly read off e.g.\ the values of the
tracial state on the $K_0$-group.

We close this introduction with the remark that
the mathematical methods used here are partly
similar to those in the theory of super selection sectors of
$2$-dimensional quantum field theory.
This seems to be due to the fact we mentioned above, namely
that the ordered $K_0$-groups of $AF$-algebras
appearing in that theory, see \cite{FSR,Boe,GoSi}, have a lot in common with
the aperiodic tilings discussed here. Moreover an approach to the fusion
structure by means of these ordered $K_0$-groups has been obtained in
\cite{Rec}. For example
Chebyshev polynomials of the second kind (which reflect the fusion structure
of $SU(2)$-WZW-models) naturally appear in the analysis of the
structure of the tilings.

\section{Quantum Spaces Related To $ADE$-Root Systems}

\subsection{Classes of Locally Isomorphic Tilings}

Classes of locally isomorphic aperiodic tilings (modulo translation)
yield examples for
- following the terminology of A. Connes - quantum spaces. Their definition
is content of this section.
\bigskip

\noindent The aperiodic tilings considered here can be obtained by
the strip method, a detailed description of which is given in
\cite{DuKa} or, for the
special case of the Penrose tilings, in \cite{Bru}. We shall briefly
describe some of their results.
A set of aperiodic tilings may be defined with the help of
\begin{itemize}
\item an $N$-dimensional abelian lattice $\Lambda \subset \Real^{N}$,

\item an $n$-dimensional subspace $P \subset \Real^{N}$,

\item a chosen basis $\cal L$ of $\Lambda$.
\end{itemize}
Given $\cal L$ the subset
\begin{equation}                                 \label{n1}
\gamma = \{\sum_{\alpha\in{\cal L}}\lambda_{\alpha}\alpha| \lambda
\in (0,1)\}
\end{equation}
of $\Real^N$ (which we also denote by $\Real\Lambda$) is the interior
of a fundamental domain of $\Lambda$.
For any $a\in P^{\perp}$, the (Euclidian) orthogonal
complement of $P$, one may construct the {\em strip} along $P$
\begin{equation}                                 \label{n2}
S_{\gamma-a} = \{x+y-a \mid x\in \gamma , y\in P\} ,
\end{equation}
which is an open subset of $\Real^N$.
Its intersection with the lattice $\Lambda$ is orthogonally
projected onto $P$ furnishing the points of an
$n$-dimensional tiling, which shall be denoted by $T_{\gamma-a}$.
We denote by $\pi$ resp.\ $\pi^{\perp }$ the orthogonal projection onto
$P$ resp.\ $P^{\perp }$, hence
\begin{equation}                                 \label{n3}
T_{\gamma-a} = \pi (S_{\gamma-a}\cap\Lambda ).
\end{equation}
The set $\cal L$ indicates which of the lattice points are joined
by a link, namely all pairs of the form ($\lambda,\lambda+\alpha$),
$\lambda\in\Lambda$, $\alpha\in{\cal L}$.
The links of the tiling are then given by the projection of all links
in $\Real^N$ which lie completely in $S_{\gamma-a}$.
We denote by ${\cal T}_{\gamma-a}$ the whole tiling, i.e.\ the vertex set
$T_{\gamma-a}$ together with the links.
\bigskip

\noindent
The best known examples of $2$-dimensional tilings are given by the following
projections to which we refer as $\Integer^N$-projections.
The lattice $\Lambda = \Integer^N$ with standard basis
${\cal L}=\{e_1,\cdots,e_N\}$
is invariant under the $N$-fold symmetry
$\omega : e_i\mapsto e_{i+1}$, ($e_{N+1}=e_1$). This symmetry furnishes a
decomposition of $\Real^N$ into subspaces which are stable under its action
\[\Real^N = \bigoplus_{0\leq m \leq \frac{N}{2}}  P^{(m)} ,\]
$m$ being an integer. Hereby $\omega$ acts on the planes $P^{(m)}$,
$0 < m < \frac{N}{2}$, as a rotation around $\frac{2\pi m}{N}$,
on $P^{(0)}$, which is the symmetry axis, as the identity, and,
if $N$ is even, on $P^{(\frac{N}{2})}$ as a reflection.
Now one takes $P=P^{(1)}$.
In particular for $N=5$ in this way all Penrose tilings are obtained as
${\cal T}_{\gamma-a}$ with $a\in P^{(2)}$.
\bigskip

\noindent
A tiling is called aperiodic, if there is no translation $y\in P$ such
that ${\cal T}_{\gamma-a}-y={\cal T}_{\gamma-a}$.
In the cases being investigated in this work, in which $\Lambda$ is either
$\Integer^n$ or a root lattice, one may establish the following
results \cite{DuKa}:
\begin{itemize}
\item
${\cal T}_{\gamma-a}$ is aperiodic, iff $P\cap \Lambda = \{\vec{0}\}$.
\item
$\pi^{\perp }(\Lambda )$ is dense
in $P^{\perp }$, iff $P^{\perp }\cap \Lambda = \{\vec{0}\}$.
\end{itemize}
\smallskip

\noindent
Comparing different values for the parameter $a$ yields the following:
\begin{enumerate}
\item
If $a'-a =\pi^{\perp}(\lambda )$ for some
$\lambda\in\Lambda$, then $T_{\gamma-a} = T_{\gamma-a'}-\pi(\lambda)$,
and hence ${\cal T}_{\gamma-a}$ and ${\cal T}_{\gamma-a'}$
differ by an overall translation on $P$. It is therefore natural to
identify them.
\item
If $a'-a\in\overline{\pi^{\perp}(\Lambda)}$ (the closure
of $\pi^{\perp}(\Lambda)$), then for any finite subset
$X\subset T_{\gamma-a}$
there are infinitely many translations $\pi(\lambda)$ on $P$ such that
$X \subset T_{\gamma-a'}-\pi(\lambda)$.
\item
If however $a'-a\notin\overline{\pi^{\perp}(\Lambda)}$, then in general
there are finite patterns in ${\cal T}_{\gamma-a}$ which do not occur in
${\cal T}_{\gamma-a'}$.
\end{enumerate}
Due to the identification of tilings through overall
translations on $P$ one is tempted to say that the elements of
\begin{equation}                        \label{sin1}
\{{\cal T}_{\gamma-a} \mid a \in
\overline{\pi^{\perp}(\Lambda)}\}\hbox{\ mod translation}  ,
\end{equation}
are locally indistinguishable.
However, there is a subtle complication
arising from the fact that this set contains singular tilings.
To illustrate the notion of a singular tiling
- which is the analog of a singular grid in de
Bruijn's approach \cite{Bru} - let us for the moment simplify the
situation by considering $\Lambda = \Integer^n$ and $P$ being a
$1$-dimensional subspace spanned by a vector having positive entries
with respect to the basis ${\cal L}$ such that
$P^{\perp}\cap\Lambda=\{\vec{0}\}$.
Then for almost all $a \in
\overline{\pi^{\perp}(\Lambda)}=P^{\perp}$
the points $T_{\gamma-a}$ divide $P$
completely into intervals of length $\pi(\alpha)$,
$\alpha\in {\cal L}$, these intervals being the prototiles.
Such a tiling shall be called regular.
However, care has to be taken if the boundary
$\partial S_{\gamma-a}$ of $S_{\gamma-a}$ has nonempty intersection with
$\Lambda$. If $v\in\partial S_{\gamma-a}\cap\Lambda$,
then by construction there is a unique
$\alpha\in{\cal L}$ such that $v-\alpha\in S_{\gamma-a}\cap\Lambda$, but no
$\alpha'\in{\cal L}$ satisfies $v-\alpha+\alpha'\in S_{\gamma-a}$.
Hence $\pi(v-\alpha)\in T_{\gamma-a}$ is the boundary point of an interval
which is not of length $\pi(\alpha')$ for some $\alpha'\in{\cal L}$.
In fact at least
one point is missing to make it a prototile; one says the tiling
has a hole.
Such a tiling is certainly
locally distiguishable from regular
tilings and is consequently called singular.

In general a regular tiling may be defined as follows:
A tiling ${\cal T}_{\gamma-a}$ shall
be called {\em regular} if for any bounded subset $U\subset P$ there
is an open neighborhood $V\subset\overline{\pi^{\perp}(\Lambda)}$
of $\vec{0}$ such that for all $a'-a\in V$:
$\pi^{-1}(U)\cap S_{\gamma-a}\cap\Lambda =
\pi^{-1}(U)\cap S_{\gamma-a'}\cap\Lambda$,
$\pi^{-1}(U)$ denoting the preimage of $U$.
In other words, a regular tiling is locally stable
against small perturbations of $a$ inside $\overline{\pi^{\perp}(\Lambda)}$.
Non-regular tilings are also called singular tilings.
We denote $K_{\gamma} = \{a\in \pi^{\perp}(\gamma)\cap
\overline{\pi^{\perp}(\Lambda)}\,|\,{\cal T}_{\gamma-a}\hbox{\ regular}\}$
\begin{lem}                                      \label{ll1}
Assume $\overline{\pi^{\perp}(\Lambda)}=P^{\perp}$.
Then ${\cal T}_{\gamma-a}$ is regular
iff $\partial S_{\gamma-a}\cap \Lambda = \emptyset$.
\end{lem}
{\em Proof:}
Let $\partial S_{\gamma-a}\cap \Lambda = \emptyset$
 and $U\subset P$ bounded.
Then there
is an open neighborhood $V\subset P^{\perp}$ of $\vec{0}$
such that
$\pi^{-1}(U)\cap\{\partial S_{\gamma-a} - b\}\cap\Lambda =\emptyset$
for all $b\in V$.
Hence $\pi^{-1}(U)\cap S_{\gamma-a-b}\cap\Lambda =
\pi^{-1}(U)\cap S_{\gamma-a}\cap\Lambda$
and therefore ${\cal T}_{\gamma-a}$ is regular.
Now assume $v\in\partial S_{\gamma-a}\cap \Lambda$, in particular
$v\notin S_{\gamma-a}$. There ought to be a
$v_0\in P^{\perp}$ such that $v-c v_0\in S_{\gamma-a}$ for all
$c\in (0,1)$, i.e $v\in S_{\gamma-a+c v_0}\cap\Lambda$. As any open
neighborhood $V\subset P^{\perp}$ of $\vec{0}$ has nonzero intersection with
$\bigcup_{c\in (0,1)}\{ c v_0\}$,
${\cal T}_{\gamma-a}$ can not be regular. \hfill $\Box$
\bigskip

\noindent
{\em Remark:} The above lemma shows that in the case of
$\overline{\pi^{\perp}(\Lambda)}=P^{\perp}$
regular tilings do not have holes. If this condition is not satisfied,
holes may appear even in regular tilings, but then they occur periodically.
\bigskip

\noindent
Translation of a regular tiling yields a regular one so that
we may restrict (\ref{sin1}) to regular tilings hereby obtaining
tilings that are indeed locally indistinguishable:
\begin{prop}
Let ${\cal T}_{\gamma-a}$ and ${\cal T}_{\gamma-a'}$ be regular tilings with
$a,a'\in\overline{\pi^{\perp}(\Lambda)}$.
For any bounded subset $U\subset P$
there are infinitely many translations $\pi(\lambda)$ on $P$,
$\lambda\in\Lambda$, such that
$U\cap {\cal T}_{\gamma-a}=U\cap \{{\cal T}_{\gamma-a'}-\pi(\lambda)\}$.
\end{prop}
{\em Proof:}
Let $U\subset P$ be bounded and $V\subset\overline{\pi^{\perp}(\Lambda)}$
be an open neighborhood of $\vec{0}$ such that
$\pi^{-1}(U)\cap S_{\gamma-a}\cap\Lambda =
\pi^{-1}(U)\cap S_{\gamma-a-b}\cap\Lambda$ for all
$b\in V$.
As $a'-a\in\overline{\pi^{\perp}(\Lambda)}$ there are infinitely many
$\lambda\in\Lambda$ such that
$a-a'-\pi^{\perp}(\lambda)\in V$.
Taking these values for $b$ showes
$U\cap {\cal T}_{\gamma-a}=U\cap \{{\cal T}_{\gamma-a'}-\pi(\lambda)\}$.
\hfill$\Box$
\bigskip

\noindent
Two tilings satisfying the property that every finite pattern of the one
occurs in the other (and vice versa) are called
{\em locally isomorphic} \cite{GrSh}. This motivates the consideration of
so called $LI$-classes \cite{LeSt,SoSt} of tilings, a concept
which is designed for any tiling of $P$ by prototiles. The $LI$-class
represented by a tiling ${\cal T}_{\gamma-a}$ for $a\in K_{\gamma}$
is given by all tilings ${\cal T}$ of $P$
which are locally isomorphic to ${\cal T}_{\gamma-a}$ meaning that
for any bounded $U\subset P$
there is a translation $y$ on $P$, such that
$U\cap {\cal T}=U\cap \{{\cal T}_{\gamma-a}-y\}$.
By the above proposition this $LI$-class contains all ${\cal T}_{\gamma-a'}$
with $a' \in K_{\gamma}$ (values $a'\notin\overline{\pi^{\perp}(\Lambda)}$
will in general lead to different $LI$-class).
Now one considers the quotient of the $LI$-class represented by the tilings
with
$a \in K_{\gamma}$ modulo translation. It is convenient
to restrict the total space to
\begin{equation}
\Omega = \{ {\cal T}\,|\,\forall U\subset P\hbox{ bounded } \exists
a\in K_{\gamma}:
U\cap {\cal T} = U\cap {\cal T}_{\gamma-a}\} ,
\end{equation}
for we shall see in section~2.1 that $\Omega$ can be equipped with a topology
with respect to which it is totally disconnected and compact.
The above quotient then being expressed as
\begin{equation}
\Psi = \Omega / \hbox{transl.}
\end{equation}
furnishes for aperiodic tilings an example of a
{\em quantum space} as described in A. Connes' book \cite{Con}.
This particular name originates in the failure of classical
(commutative) geometry to describe it properly.
In fact as $\Psi$ is in that case
the quotient of a topological space modulo a
dense equivalence relation its quotient topology is not Hausdorff,
namely the only closed sets are $\Psi$ itself and the empty set, i.e.\
it appears, analyzed
in terms of classical topology, as a single point \cite{Con}. However
its structure is much richer.
It may be revealed by assigning a certain non-commutative $C^*$-algebra
to $\Psi$ and investigating its invariants.
This program has been carried out in \cite{Con} for the Penrose tilings.
The final goal of this work shall be a generalization of it to classes of
tilings which are defined by $ADE$-root systems.

\subsection{Tilings Defined by Root Systems}      \label{ss1}

We shall apply the strip method to the root lattice
$\Lambda$ of a root system. The subspace $P$ along which the strips
are constructed
is characterized by its invariance under the action of a
Coxeter element of the Weyl group of reflections in $\Lambda$.
\bigskip

\noindent Any root $\alpha$ of the root system defines an automorphism
of the root lattice
$s_{\alpha} : \Lambda \mapsto \Lambda$
\begin{equation}
s_{\alpha}(\beta) := \alpha - 2\frac{(\alpha,\beta)}{(\alpha_,\alpha)}
\beta
\end{equation}
which may be considered as a reflection in the Euclidian space $\Real^N$,
$(.,.)$ denoting its scalar product, in which $\Lambda$ is embedded.
Having choosen a system of simple (or fundamental) roots $\{\alpha_i\}_i$,
the group
generated by the elements $s_i=s_{\alpha_i}$ contains all the above
reflections; it is called Weyl group of
reflections. It is a Coxeter group, i.e.\ the relations of its generators
are of the form
\begin{equation}
(s_{i}s_{j})^{m_{ij}} = 1,                              \label{cox1}
\end{equation}
where $m_{ij} = m_{ji} \in \Integer^{\geq 2}$ for $i\neq j$ and $m_{ii} = 1$.
In particular
all generators are of order 2 and $s_{i},s_{j}$ commute, iff $m_{ij}=2$.
The relations (\ref{cox1})
may be uniquely encoded in a so called Coxeter graph
$\Gamma$ consisting of a set of vertices $\Gamma^{(0)}$ and a set of
unoriented edges $\Gamma^{(1)}$.
The vertices are associated to the simple roots,
$\Gamma^{(0)} \cong \{\alpha_{1},\alpha_{2},..\}$,
the $i$'th and $j$'th vertex are connected by
\begin{equation}
{\cal C}_i^j:=m_{ij}-2
\end{equation}
edges, and
no edge has equal source and range: ${\cal C}_i^i=0$.
The matrix
$\cal C$ is called connectivity (or adjacency)
matrix of the graph.
A particular role in the analysis of Coxeter groups is played by
elements which are products of all generators. They are called
Coxeter elements.
Two distinct Coxeter elements $\omega$, $\tilde{\omega}$
differ by the succession
of the generators in the product, but they are conjugate, i.e.\ there is a
Weyl-reflection $\delta$, such that
\begin{equation}                                  \label{cox2}
\tilde{\omega} = \delta\omega\delta^{-1}.
\end{equation}
In particular all Coxeter elements have the same order $h$ ($\omega^h=1$),
called the Coxeter number of the group, and the same eigenvalues.
The eigenvalues, coming, if complex,
in conjugate pairs, are of the form $exp(\frac{2\pi im}{h})$,
the possible values for $m\in \Integer_h$ being called Coxeter exponents
here denoted by $Exp(\Gamma)$.
$0$ is never but $1$ is always a Coxeter exponent.
$\Real^N$ may now be decomposed into invariant subspaces
of $\omega$
\begin{equation}
\Real^N = \bigoplus_{m\in Exp(\Gamma)^{\leq\frac{h}{2}}} P_{\omega}^{(m)},
\end{equation}
$P_{\omega}^{(m)}$ denoting the subspace on which $\omega$ acts as a
rotation around $\frac{2\pi m}{h}$ or, if
$m=\frac{h}{2}$, as a reflection. Consequently the direct sum runs over
roughly speaking half of the Coxeter exponents (one for any complex pair
of eigenvalues)
eventually including $\frac{h}{2}$.
\bigskip

\noindent
Given a root system we now consider
$2$-dimensional
tilings which are defined with the help of
\begin{itemize}
\item the root lattice $\Lambda$,
\item $P_{\omega} = P_{\omega}^{(1)}$ being specified by the choice of
a Coxeter element $\omega$,
\item a base ${\cal L}$ of $\Lambda$.
 \end{itemize}
How does the choice of a Coxeter element affect that construction?
Whereas the set $Exp(\Gamma)$,
is independent of the choice of $\omega$,
the decomposition of $\Real^N$ into invariant subspaces
$P_{\omega}^{(m)}$ is not.
If two Coxeter elements
$\omega$, $\tilde{\omega}$ are related by $\delta$ as in (\ref{cox2}),
then
\begin{equation}                                          \label{cox13}
P_{\tilde{\omega}}^{(m)} =
\delta\,P_{\omega}^{(m)}
\end{equation}
and in particular
$\tilde{\pi} = \delta\pi\delta^{-1}$, $\tilde{\pi}$ resp.\ $\pi$ denoting
the projection onto $P_{\tilde{\omega}}^{(1)}$ resp.\ $P_{\omega}^{(1)}$.
Therefore
\begin{equation}
\tilde{S}_{\gamma} = \{\gamma+P_{\tilde{\omega}}\}
  =  \delta\,\{\delta^{-1}\gamma+P_{\omega}\} =
\delta\,S_{\delta^{-1}\gamma}
\end{equation}
as well as
\begin{equation}
\tilde{T}_{\gamma} =
\tilde{\pi}(\tilde{S}_{\gamma}\cap\Lambda)
  =  \delta\pi\delta^{-1}
(\tilde{S}_{\gamma}\cap\Lambda) =
\delta\,T_{\delta^{-1}\gamma}.
\end{equation}
Moreover if $(\lambda,\lambda+\alpha)\in
\tilde{S}_{\gamma}$, i.e.\ $\tilde{\pi}(\lambda,\lambda+\alpha)$
is a link on
$P_{\tilde{\omega}}$, then
$\delta^{-1}(\lambda,\lambda+\alpha)\in
S_{\delta^{-1}\gamma}$, i.e.\ $\pi \delta^{-1}(\lambda,\lambda+\alpha)$
is a link on
$P_{\omega}$.
Hence up to $\delta$, which acts as an isometry on $P$,
a change of the Coxeter element by
$\omega \mapsto \delta\omega\delta^{-1}$ has the same effect as
transforming the links by
${\cal L} \mapsto \delta^{-1}{\cal L}$
(which amounts to $\gamma\mapsto\delta^{-1}\gamma$).
We may therefore concentrate on a particular choice for the Coxeter element.

\subsection{Decomposition into $1$-Dimensional Tilings}   \label{sub1}

For further discussions we have to restrict ourself to root systems of
type
$ADE$, whose Coxeter graphs are simply laced and bipartite (bicoloured).
A Coxeter
graph $\Gamma$ is called bipartite, if the set of vertices $\Gamma^{(0)}$
decomposes into a disjoint union of two subsets
$\Gamma^{(0)}_{\epsilon}$, $\epsilon\in \Integer_2$,
such that never two elements of the same
subset are linked. As a consequence for the group structure two generators
assigned to the same subset
commute, and the products
\begin{equation}
s_{(\epsilon)} = \prod_{i\in\Gamma^{(0)}_{\epsilon}}\,s_i
\end{equation}
of all generators of $\Gamma^{(0)}_{\epsilon}$ are
independent of the ordering.
The Coxeter element
\begin{equation}                               \label{b3}
\omega = s_{(1)}s_{(2)}
\end{equation}
is (up to inversion, $\omega^{-1}=s_{(2)}s_{(1)}$) canonical, since the
above decomposition of $\Gamma^{(0)}$ is unique. We simply denote
$P^{(m)} = P^{(m)}_{\omega}$, if they are determined by this canonical
$\omega$, and call
$P = P_{\omega}^{(1)}$ the
canonical plane.

On the one hand we may now decompose the root
lattice $\Lambda$ as well as $P^{(m)}$ into the sublattices
$\Lambda_{\epsilon}$, which are generated by the simple roots assigned to
$\Gamma^{(0)}_{\epsilon}$ as well as the subspaces
$P^{(m)}_{\epsilon} = P^{(m)}\cap \Real\Lambda_{\epsilon}$.
On the other hand $\Real^N$ may be decomposed into the eigenspaces
of the connectivity matrix ${\cal C}\in End(\Lambda)\subset
End(\Real^N)$. The latter may be written as
\begin{equation}
{\cal C} = \sum_{m\in Exp(\Gamma)} \tau^{(m)}\pi^{(m)},
\end{equation}
where $\pi^{(m)}$ denotes the projection onto the eigenspace $E^{(m)}$
of the eigenvalue
\begin{equation}
\tau^{(m)} = 2cos(\frac{m\pi}{h})         \label{pol5}
\end{equation}
of $\cal C$, $h$ the
Coxeter number of $\Gamma$.
These two decompositions of $\Real^N$ are related as follows:

\begin{lem}                                            \label{lem4}
For $m\neq\frac{h}{2}$:
$P^{(m)}=E^{(m)}\oplus E^{(h-m)}$.
\end{lem}
{\em Proof:}
The main part of the proof is given in the book of Carter \cite{Car}.
Concerning this part we state only some details, which are needed later on.
Remember that in the simply laced case
${\cal C}_{ij} = 2 - 2(\alpha_i,\alpha_j)$
(with normalization $(\alpha_i,\alpha_i)=1$).
Let $\vec{\nu}^{(m)}=\sum_i\nu_{\:i}^{(m)}\alpha_i$ be
an eigenvector of
$\cal C$ to the eigenvalue $\tau^{(m)}$, i.e.\
\begin{equation}                                     \label{span1}
E^{(m)} = \Real\vec{\nu}^{(m)}.
\end{equation}
Furthermore let
$\vec{\nu}^{(m)}_{\epsilon}=\sum_{j\in\Gamma_{\epsilon}^{(0)}}
\nu_{\:j}^{(m)}\alpha_j$.
Clearly $\vec{\nu}^{(m)}_{1} + \vec{\nu}^{(m)}_{2}
= \vec{\nu}^{(m)}$.
In \cite{Car}
the dual basis
$\{\hat{\alpha}_i\}_i$ is used to define
$\vec{\mu}^{(m)}_{\epsilon}=\sum_{j\in\Gamma_{\epsilon}^{(0)}}
\nu_{\:j}^{(m)}\hat{\alpha}_j$.
They are shown to have the following properties:
\begin{enumerate}
\item Both,
$\vec{\mu}^{(m)}_{1}$ and
$\vec{\mu}^{(m)}_{2}$ have the same length and form an angle of
$\frac{m\pi}{h}$.
\item The canonical Coxeter element acts on the linear span of
$\vec{\mu}^{(m)}_{1}$ and
$\vec{\mu}^{(m)}_{2}$ as rotation around $\frac{2m\pi}{h}$, and hence
this span equals $P^{(m)}$.
\end{enumerate}
Now $\alpha_i =\sum_j (\delta_{ij} -\frac{1}{2}{\cal C}_{ij})\hat{\alpha}_j$
may be used to obtain
$\vec{\nu}^{(m)}_{\epsilon} = \vec{\mu}^{(m)}_{\epsilon} -
 cos(\frac{m\pi}{h})\vec{\mu}^{(m)}_{\epsilon+1}$, which implies that
$\vec{\nu}^{(m)}_{1}$ and $\vec{\nu}^{(m)}_{2}$ form an angle of
$\frac{(h-m)\pi}{h}$ and span $P^{(m)}$, too. Therefore
\begin{equation}                                  \label{span2}
P^{(m)}_{\epsilon}=\Real\vec{\nu}_{\epsilon}^{(m)}.
\end{equation}
On the other hand the bipartition of the graph implies
${\cal C}(\Lambda_{\epsilon}) \subset \Lambda_{\epsilon+1}$ and hence
${\cal C}\vec{\nu}_{\epsilon}^{(m)} = \tau^{(m)}
\vec{\nu}_{\epsilon+1}^{(m)}$. Therefore
\[ {\cal C}(\vec{\nu}_{1}^{(m)} -
\vec{\nu}_{2}^{(m)}) = \tau^{(m)}
(\vec{\nu}_{2}^{(m)} -
\vec{\nu}_{1}^{(m)}) = \tau^{(h-m)}
(\vec{\nu}_{1}^{(m)} -
\vec{\nu}_{2}^{(m)}). \]
Hence $E^{(h-m)}$ is spanned by
$\vec{\nu}_{1}^{(m)} -
\vec{\nu}_{2}^{(m)}$.\hfill $\Box$
\bigskip

\noindent The first of the above decompositions furnishes a decomposition of
the tilings, provided the links ${\cal L}$ decompose, too:

\begin{thm}                         \label{b4}
If ${\cal L} = {\cal L}_{1}\cup {\cal L}_{2}$, where
${\cal L}_{\epsilon}\subset \Lambda_{\epsilon}$, then any
$2$-dimensional tiling related to $(\Lambda,P,{\cal L})$ decomposes into a
cartesian product of two $1$-dimensional tilings. More precisely
\begin{equation}
T_{\gamma-a}  =
T_{\gamma_{1}-a_{1}} \times
T_{\gamma_{2}-a_{2}}      \label{krz}
\end{equation}
where $\gamma_{\epsilon} = \{\sum_{\alpha\in{\cal L}_{\epsilon}}
\lambda_{\alpha}\alpha|\lambda_{\alpha}\in (0,1)\}$,
$T_{\gamma_{\epsilon}-a_{\epsilon}}$
are the points of the $1$-dimensional
tiling related to
$(\Lambda_{\epsilon}$,$P_{\epsilon},{\cal L}_{\epsilon})$ and
$a_{\epsilon}$ is obtained by decomposing $a$ into the direct sum of
vectors of $P^{\perp}_{\epsilon} = P^{\perp}\cap\Real\Lambda_{\epsilon}$.

\end{thm}
{\em Proof:}
(\ref{krz}) is to be interpreted as follows: For every $t\in T_{\gamma-a}
\subset P$ there is a unique decomposition $t=t_{1}+t_{2}$ with
$t_{\epsilon}\in T_{\gamma_{\epsilon}-a_{\epsilon}}
\subset P_{\epsilon}$,
and vice versa for every pair
$t_{\epsilon}\in T_{\gamma_{\epsilon}-a_{\epsilon}}$,
$\epsilon=1,2$,
the sum $t=t_{1}+t_{2}$ lies in $T_{\gamma-a}$.
This follows directly from the assumption (which implies that the
elements of $\gamma$ have a unique decomposition too), if the projection onto
$P$ preserves the decomposition, i.e.\
if $\pi(\Lambda_{\epsilon})\subset P_{\epsilon}$.\\
Setting $\hat{\nu}_{\epsilon} = \frac{\vec{\nu}_{\epsilon}^{(1)}}
{\|\vec{\nu}_{\epsilon}^{(1)}\|}$ we have for $a\in \Real^N$
\[\pi(a) = \frac{(\hat{\nu}_{1}+\hat{\nu}_{2},a)}{2(1-cos\frac{\pi}{h})}
(\hat{\nu}_{1}+\hat{\nu}_{2}) +
\frac{(\hat{\nu}_{1}-\hat{\nu}_{2},a)}{2(1+cos\frac{\pi}{h})}
(\hat{\nu}_{1}-\hat{\nu}_{2}).\]
Using
\[(\hat{\nu}_{\epsilon},a) =
 \sum_{i,k}\hat{\nu}_{\epsilon\,i}
(\delta_{ij}-\frac{1}{2}{\cal C}_{ij})\,a_j =
 \sum_{i\in\Gamma_{\epsilon}^{(0)}}\hat{\nu}_{\epsilon\,i}\,a_i
 - cos\frac{\pi}{h}
 \sum_{i\in\Gamma_{\epsilon+1}^{(0)}}\hat{\nu}_{\epsilon+1\,i}\,a_i\]
one obtains
\begin{equation}
\pi(a) = \sum_{i\in\Gamma_{1}^{(0)}}\hat{\nu}_{1\,i}\,a_i \,\hat{\nu}_{1}+
 \sum_{i\in\Gamma_{2}^{(0)}}\hat{\nu}_{2\,i}\,a_i \,\hat{\nu}_{2},
\end{equation}
and in particular
$\pi(\Lambda_{\epsilon})\subset P_{\epsilon}$.
As ${\cal L}_{\epsilon}\subset \Lambda_{\epsilon}$ any link of the
$2$-dimensional tiling
is either parallel to $P_{1}$ or to $P_{2}$, it is therefore
either in the $1$-dimensional tiling related to
$P_{1}$ or in the one related to $P_{2}$.\hfill $\Box$
\bigskip

\noindent
For application of this theorem we mainly consider
what we call $ADE$-projections.
These are tilings defined (next to the root lattice
and the canonical plane) by the basis of simple roots
${\cal L}=\{\alpha_i\}_i$.
The vector $\vec{\nu}^{1}_{\epsilon}$
spanning $P_{\epsilon}$ is the Perron Frobenius vector of
${\cal C}^2|_{\Lambda_{\epsilon}}$ having only strictly positive (or strictly
negative) entries.
Decomposition of a regular tiling yields regular tilings and we may
think of the latter as twosided infinite
sequences of intervals of the form
$(\pi(x),\pi(x+\alpha_i))$, at least
in case $P^{\perp}\cap\Lambda_{\epsilon}=\{\vec{0}\}$.
For $P^{\perp}\cap\Lambda_{\epsilon}\not=\{\vec{0}\}$ holes may appear even
in regular tilings, as mentioned earlier, but they occur periodically.
In order to treat both cases in the same manner
we shall fix these holes by adding some of the points that are on
$\partial S_{\gamma_{\epsilon}-a_{\epsilon}}\cap\Lambda_{\epsilon}$
periodically.

\bigskip

\noindent
{\em Remark:}
It is noteworthy that in perturbed
conformal field theories which are
related to $ADE$-Lie algebras an explanation of the allowed mass triangles
has been proposed using mathematical
methods which have a substantial overlap to the one used here
\cite{LW},\cite{Dor}:
By projection of a certain subset of three roots
(satisfying what is called fusing condition in \cite{Dor})
onto $P^{(m)}$ a reformulation of the bootstrap equations is obtained.
In this framework $\nu_i^{(m)}$ corresponds for $m>1$
to the components
of the conserved higher spin charges and for $m=1$ to the particle masses.

\subsection{Aperiodicity and Further}

This and the following subsection are devoted to the study of the
properties of the tilings related to $ADE$-root systems.
As we are interested in aperiodic tilings, the computation of
$P\cap\Lambda$ will be treated first. Hereafter
 we turn into the analysis of the structure of
$P^{\perp}\cap\Lambda$, which is of relevance for the definition of
the quantum space $\Psi$. We determine
a subspace $E$ of $\Real\Lambda$, of which may be shown that
\begin{itemize}
\item $P\subset E$ and the dimension
of $E$ resp.\ $E^{\perp}$ as vector spaces equal the dimension of
$E\cap\Lambda$ resp.\ $E^{\perp}\cap\Lambda$ as lattices.
\end{itemize}
Furthermore,
under certain circumstances
\begin{itemize}
\item $E^{\perp}\cap\Lambda = P^{\perp}\cap\Lambda$,
\end{itemize}
i.e.\ the connected component of $\overline{\pi^{\perp}(\Lambda)}$
that contains $\vec{0}$ equals $E^{\perp}$.
We may concentrate onto the invariant eigenspaces $P^{(m)}$ of the
canonical Coxeter element, since
$P_{\omega}\cap\Lambda$ resp.\ $P_{\omega}^{\perp}\cap\Lambda$ and
$P_{\tilde{\omega}}\cap\Lambda$ resp.\ $P_{\tilde{\omega}}^{\perp}
\cap\Lambda$
are related by the automorphism $\delta$ as in (\ref{cox13}).
\bigskip

\noindent Concerning the aperiodicity of the tilings we have:
\begin{prop}                                        \label{lll1}
 $P^{(m)}\cap\Lambda = \{\vec{0}\}$ iff $(\tau^{(m)})^2\notin \Rational$.
\end{prop}
{\em Proof:} Remember (\ref{span2}) and
${\cal C}(P^{(m)}_{\epsilon}) = P^{(m)}_{\epsilon+1}$. Hence
 $P^{(m)}\cap\Lambda\neq\{\vec{0}\}$ iff
 $P^{(m)}_1\cap\Lambda_1\neq\{\vec{0}\}$ iff
 $P^{(m)}_2\cap\Lambda_2\neq\{\vec{0}\}$. Moreover
 $\nu^{(m)}_{\epsilon}$ is an eigenvector of ${\cal C}^2$ to the eigenvalue
 $(\tau^{(m)})^2$.
 Now assume $\vec{0}\neq z\in
 P^{(m)}_{\epsilon}\cap\Lambda_{\epsilon}$. Then
 $(\tau^{(m)})^2 z =
 {\cal C}^2 z\in
 P^{(m)}_{\epsilon}\cap\Lambda_{\epsilon}$ implying
 $(\tau^{(m)})^2\in \Rational$.
Conversely let
 $(\tau^{(m)})^2\in \Rational$ and $a\in \Integer$ such that
 $a(\tau^{(m)})^2\in \Integer$. This implies
$\{\vec{0}\}\neq Ker_{\Integer}(a({\cal C}^2 - (\tau^{(m)})^2) \subset
 P^{(m)}\cap\Lambda$.\hfill $\Box$
 \bigskip

\noindent As rationality of $(\tau^{(1)})^2$ does only occur
for $h=2,3,4,6$, all $ADE$-root systems
except $A_2,A_3,A_5$ and $D_4$ lead to aperiodic tilings.
\bigskip

\noindent To analyse the structure of $P^{\perp}\cap\Lambda$
we use the Chebyshev polynomials of the second kind:
Consider the set of polynomials in one variable over $\Integer$,
$p_1(x) = 1$ and
\begin{equation}
p_k(x)  =  \overbrace{\left| \begin{array}{cccc}
x & 1 & 0 & \cdots    \\
1 & x & 1 & \ddots    \\
0 & 1 & x & \ddots          \\
\vdots & \ddots & \ddots & \ddots     \end{array} \right|}^{k-1}
\end{equation}
for $k\geq 2$. In fact $p_k(-x)$ is the
characteristic
polynomial of the connectivity matrix of $A_{k-1}$, the latter being
denoted
by ${\cal A}_k$.
The polynomials satisfy the recursion relations
\begin{equation}
p_{k+2}(x) = x p_{k+1}(x) - p_k(x)                         \label{pol1}
\end{equation}
and have homogeneous $\Integer_2$-degree
\begin{equation}
p_k(-x) = (-1)^{k-1}p_k(x) .
\end{equation}
For $|x|\leq 1$, $p_k(x) = U_{k-1}(\frac{x}{2})$, where $U_k(x)$
is the k'th Chebyshev polynomial of the second kind \cite{NiUv} and
\begin{equation}
U_{k-1}(cos\Theta) = \frac{sin k\Theta}{sin\Theta}.    \label{pol6}
\end{equation}
As long as $i+1\leq h-1$ an equivalent way of writing (\ref{pol1}) is
\begin{equation}
x p_i(x) = ({\cal A}_h)_i^j\, p_j(x) ,
\label{pol2} \end{equation}
which indicates that the vector $\vec{p}(x) = (p_1(x),p_2(x),\cdots,
p_{h-1}(x))$ is an eigenvector of ${\cal A}_h$ whenever $x$ is an
eigenvalue,
i.e.\ $p_h(x) = 0$. Hence for $m\in Exp(A_{h-1})$
the polynomials obey
\begin{equation}
p_i(\tau^{(m)})\, p_k(\tau^{(m)})
    = (p_i({\cal A}_h))_k^l\, p_l(\tau^{(m)})              \label{pol3}
\end{equation}
(for any $i,k\leq h-1$) and
\begin{equation}
p_k(\tau^{(m)})=
\frac{sin\frac{km\pi}{h}}{sin\frac{m\pi}{h}}.     \label{pol7}
\end{equation}
Note that $\vec{p}(\tau^{(m)})$ yields an explicit realization of the
vector spanning $E^{(m)}$ in the case of the Coxeter group $A_{h-1}$.
As $|p_k(\tau^{(m)})|=|p_{k+h}(\tau^{(m)})|$ it is useful to regard
the set of Coxeter exponents $Exp(\Gamma)$
as a subset of $\Integer_h$
($=\Integer\:mod\:h$), $h$ being the Coxeter number of $\Gamma$.
It may be checked that the subset of $Exp(\Gamma)$ of elements
which are invertible
in $\Integer_h$ does only depend on the Coxeter number, in fact
\begin{equation}
 \{m\in Exp(\Gamma)\mid m\: \hbox{is invertible in}\,
  \Integer_h\} = \Integer^*_h ,
\end{equation}
$\Integer^*_h$ denoting the group of invertible elements
of $\Integer_h$. Furthermore as long as $k\in\Integer^*_h$ one has
$k\,Exp(\Gamma) = Exp(\Gamma)$ and $k\frac{h}{2} = \frac{h}{2}$,
if $\frac{h}{2} \in Exp(\Gamma)$.

\begin{lem}                                      \label{lem3}
Let $\Gamma$ be an $ADE$-Coxeter group with Coxeter number $h$,
$\cal C$ the
connectivity matrix of its graph.
If $k\in \Integer^*_h$, then $p_k({\cal C})$ is an automorphism of $\Lambda$,
whereas otherwise $p_k({\cal C})$ is not even injective.
\end{lem}
{\em Proof:}
At first let $k\in \Integer^*_h$.
We have
\[ \prod_{m\in Exp(\Gamma)}| sin(\frac{m\pi}{h})| =
 \prod_{m\in k\,Exp(\Gamma)}| sin(\frac{m\pi}{h})| =
 \prod_{m\in Exp(\Gamma)}| sin(\frac{km\pi}{h})| \]
which, in view of (\ref{pol7}), implies
\begin{equation}
\prod_{m\in Exp(\Gamma)} | p_k(\tau^{(m)})| = 1.
\end{equation}
But this implies for the determinant of $p_k({\cal C})$ to be of modulus $1$
(if $\frac{h}{2}\in Exp(\Gamma)$, then $|p_k(\tau^{(\frac{h}{2})})| = 1$,
hence the statement is also true for $D_n$ with $n$ even) proving
the first part of the lemma. The second one follows from
(\ref{pol7}) which implies
$| p_k(\tau^{(m)})| =0$ iff $km=0\:{\em mod}\:h$, and the
determinant of $p_k({\cal C})$ vanishes for
$k\notin \Integer_h^*$.\hfill $\Box$
\bigskip

\noindent Define
\begin{equation}
 E = \bigoplus_{m\in \Integer^*_h} E^{(m)},
\end{equation}
and $E^{\perp}$, its orthogonal complement in $\Real^N$.
Clearly
$P\subset E$ and $E^{\perp}\subset P^{\perp}$.
\begin{prop}                                      \label{lem2}
$E\cap\Lambda$ and
$E^{\perp}\cap\Lambda$ have both maximal dimension, i.e.\
the dimension of $E\cap\Lambda$ resp.\
$E^{\perp}\cap\Lambda$ as a lattices equal the
dimension of $E$ resp.\ $E^{\perp}$ as vector spaces.
\end{prop}
{\em Proof:}
Given any map $\phi\in End(\Lambda)\subset End(\Real^N)$,
$Im\:\phi \cap\Lambda$ as well as $Ker\:\phi \cap\Lambda$
have always maximal dimension.
Set $\phi := \prod_{k\in \Integer_h\backslash \{0\}}
 p_k({\cal C})\in End(\Lambda)$.
Viewed as an element of $End(\Real^N)$,
\[Im\:\phi = E\]
and
\[Ker\:\phi = E^{\perp},\]
following again from $| p_k(\tau^{(m)})| =0$ iff $km=0\:{\em mod}\:h$.
\hfill$\Box$
 \bigskip

 \noindent
The remaining question is,
whether $E\cap P^{\perp}\cap\Lambda=\{\vec{0}\}$ or
not.
As well as any automorphism of $\Lambda$ which preserves
$E\cap P^{\perp}$ and its orthogonal complement,
$p_k({\cal C})$ restricts, for $k\in \Integer^*_h$, to an automorphism
of $\Delta'=E\cap P^{\perp}\cap\Lambda$.
Hence $\Real\Delta'$ may be decomposed into eigenspaces of
$\cal C$, i.e.\ there is a subset $S\subset \Integer^*_h\backslash\{1,-1\}$
such that
\begin{equation}
\Real\Delta' = \bigoplus_{m\in S} E^{(m)}.
\end{equation}
Moreover $S=-S$, because of the fact that every element of $\Delta'$
decomposes
into a sum of elements of $\Lambda_{\epsilon}$ and
${\cal C}(P^{(m)}_{\epsilon}) = P^{(m)}_{\epsilon+1}$.
Since $p_k({\cal C})$ restricts for all $k\in \Integer^*_h$
to an automorphism on $\Delta'$
\begin{equation}                            \label{Ba1}
\prod_{m\in S} | p_k(\tau^{(m)})|  = 1 ,
\end{equation}
which is
\begin{equation}                            \label{Ba2}
 \prod_{m\in S}| sin(\frac{m\pi}{h})| =
 \prod_{m\in S}| sin(\frac{km\pi}{h})| .
\end{equation}
Formula (\ref{Ba2}) may be regarded as a necessary
condition for the existence of $S$. By the construction of $S$, which
in particular rules out $S=\Integer^*_h$,
it seems to be rather difficult to find such a set $S$. Let us already
mention
here that the existence of a solution of (\ref{Ba3}) (see below)
contradicts (\ref{Ba1}). In that case
$\Delta'=\{\vec{0}\}$ and $P^{\perp}\cap\Lambda = E^{\perp}\cap\Lambda$.

\subsection{Selfsimilarity}

Many aperiodic tilings furnish selfsimilar sets.
Selfsimilarity, being manifested by an inflation / deflation procedure,
is the basis of an alternative
description of the quantum space which shall be discussed in the
second part of this work.
\bigskip

\noindent
Considering tilings related to arbitrary
$(\Lambda,P,{\cal L})$ a map $\sigma\in
Aut(\Lambda)$, acting on the links by ${\cal L}\mapsto \sigma({\cal L})$
and correspondingly on the vertices
of the tiling by $T_{\gamma-a}\mapsto T_{\sigma(\gamma-a)}$, is
called an {\em inflation} for ${\cal T}_{\gamma-a}$, if
there is a real number $\tau>1$ such that \cite{Gae}
\begin{eqnarray}
T_{\sigma(\gamma-a)} & \subset & T_{\gamma-a}     \label{inf1}   \\
{\cal T}_{\sigma(\gamma-a)} & = & \tau {\cal T}_{\gamma-a} .     \label{inf2}
\end{eqnarray}
Hence the vertex set of a tiling which
admits an inflation is a selfsimilar
set, see \cite{DuKa} for the
 the possibility of selfsimilarity up to symmetries of the lattice
(e.g.\ reflections). The inverse of an inflation is a deflation.

Note that the above conditions are stated for a single tiling. An
 inflation for ${\cal T}_{\gamma-a}$ will in general not be at the
same time an inflation for ${\cal T}_{\gamma-a'}$. Nevertheless
${\cal T}_{\gamma-a'}$ would for $a'-a\in\overline{\Lambda}$
also be selfsimilar.

Specifying now to tilings related to $ADE$-root systems
that are cartesian products of $1$-dimensional tilings
(i.e.\ ${\cal L}={\cal L}_1\cup {\cal L}_2$),
we require that the inflation decomposes, too,
$\sigma = \sigma_1 \oplus \sigma_2$ where
$\sigma_{\epsilon}= \sigma|_{\Lambda_{\epsilon}}$ is now an inflation for
${\cal T}_{\gamma_{\epsilon}-a_{\epsilon}}$.
\bigskip

\noindent
The important property of an inflation is given by:

\begin{prop}     \label{prop1}
Let ${\cal L}=\{\alpha_i\}_i$ and
$\sigma$ be an inflation for some
${\cal T}_{\gamma-a}$ with decomposition as above.
Then $\sigma_{\epsilon}$ is a positive endomorphism (i.e.\ has entries
in $\Integer^+$) preserving $P_{\epsilon}$ and having $\tau$ as largest and
nondegenerate eigenvalue.
\end{prop}
{\em Proof:}
Consider a point
$x\in S_{\sigma^n_{\epsilon}(\gamma_{\epsilon}-a_{\epsilon})}
\cap\Lambda_{\epsilon}$
such that
$x'=x+\sigma^n_{\epsilon}(\alpha_i)
\in S_{\sigma^n_{\epsilon}(\gamma_{\epsilon}-a_{\epsilon})}
\cap\Lambda_{\epsilon}$
for some $\alpha_i\in {\cal L}_{\epsilon}$, $n\geq 1$. Hence
$\pi(x)$ and $\pi(x')$ are the boundary points
of a prototile in
${\cal T}_{\sigma^n_{\epsilon}(\gamma_{\epsilon}-a_{\epsilon})}$.
As the latter appears as a twosided sequence of prototiles
$n$-fold use of (\ref{inf1})
implies that there is a sequence of
$k=\sum_l (\sigma^n_{\epsilon})_{li}$ generators
$\alpha_{i_1},\cdots,\alpha_{i_k}$, such that
\begin{equation}
x'-x =                                          \label{seq2}
\alpha_{i_1}+\cdots+\alpha_{i_k},
\end{equation}
and for all $1\leq j<k$
\begin{equation}
x+\alpha_{i_1}+\cdots+\alpha_{i_j}\in S_{\gamma_{\epsilon}-a_{\epsilon}}
\backslash
S_{\sigma^n_{\epsilon}(\gamma_{\epsilon}-a_{\epsilon})} .
                                                \label{seq3}
\end{equation}
Since the r.h.s.\  of (\ref{seq2}) contains only $+$-signs
$(\sigma_{\epsilon})_{li}$ is positive.
Now remember that the vector $\vec{\nu}^{(1)}_{\epsilon}$
spanning $P_{\epsilon}$ has strictly positive entries.
Therefore, and because of $\pi_{\epsilon}^{\perp}(x'-x)$ being bounded
from above, all $\alpha_i\in{\cal L}_{\epsilon}$ have
to appear on the r.h.s.\ of (\ref{seq2}), if $n$ is large enough.
Hence, for $n$ large enough $\sigma^n_{\epsilon}$ has strictly
positive entries,
and the Perron Frobenius theorem may be applied to establish that
$\sigma_{\epsilon}$ has a nondegenerate
eigenvalue which exceeds all other eigenvalues in modulus. The corresponding
eigenvector is called Perron Frobenius vector and has only strictly positive
(or strictly negative) components. The Perron Frobenius theorem also applies
to $\sigma_{\epsilon}^\dagger$.
Now by (\ref{inf2})
$(\sigma_{\epsilon}^\dagger\vec{\nu},\alpha_i) =
 (\vec{\nu},\sigma_{\epsilon}\alpha_i) =
 \tau(\vec{\nu},\alpha_i)$ for all $i$
and therefore $\vec{\nu}$ is the Perron Frobenius eigenvector of
$\sigma_{\epsilon}^\dagger$ to the eigenvalue $\tau$. In particular
$\tau$ is the largest eigenvalue not only of
$\sigma_{\epsilon}^\dagger$ but also of
$\sigma_{\epsilon}$.
Finally observe that (\ref{inf2}) implies $\sigma_{\epsilon}\vec{\nu}
 = \tau\vec{\nu} + \vec{\mu}$ for some $\vec{\mu}\in P^{\perp}_{\epsilon}$
which is compatible with
the above only for $\vec{\mu} = \vec{0}$. \hfill $\Box$
\bigskip

\noindent
The above proposition asserts that the specific form of the inflation
will not be of importance for our investigations in the last section,
since there only the Perron Frobenius vector will be of importance.
So we are left with the question of whether all tilings
defined by root lattices allow for an inflation.
For a general treatement of selfsimilarlarity in two dimensions we
refer to \cite{Nii}.
But let us comment on a possible way to find one in our framework,
in case $P^{\perp}\cap\Lambda=E^{\perp}\cap\Lambda$.

Similar to the common method to obtain an inflation we search for an
automorphism
$\sigma'$ of $\Lambda$,
which rescales the vectors of $P$ by a factor $>1$,
preserves $P^{\perp}$ and is strictly
contracting on $E\cap P^{\perp}$ \cite{Gae} \cite{DuKa}.
In fact, under suitable circumstances
the requirements formulated above are sufficient, for the following reason:
First observe that for $\lambda\in \Delta=E^{\perp}\cap\Lambda$
\begin{equation}                                       \label{self0}
\pi(S_{\gamma-a-\lambda}\cap\Lambda) = \pi(S_{\gamma-a}\cap\Lambda).
\end{equation}
Maximalitiy of the dimension of
$\Delta$ implies
\begin{equation}                                       \label{self1}
\pi(S_{\gamma-a}\cap\Lambda) =
\sum_{b\in\pi_{E^{\perp}}(\Lambda)\cap\pi_{E^{\perp}}(\gamma-a)}
\pi(S_{\gamma-a}\cap\{E+b\}\cap\Lambda),
\end{equation}
the sum being finite,
where $\pi_{E^{\perp}}$ denotes the orthogonal projection
onto $E^{\perp}$.
Now (\ref{inf1}) would be satisfied, if for all
$b\in\pi_{E^{\perp}}(\Lambda)\cap\pi_{E^{\perp}}(\gamma-a)$ there is a
$\lambda(b)\in\Delta$ such that
\begin{equation}                                       \label{self2}
\sigma(S_{\gamma-a}\cap\{E+b\})-\lambda(b) \subset
S_{\gamma-a}\cap\{E+\sigma(b)-\lambda(b)\} .
\end{equation}
As $\pi_{E^{\perp}}(\gamma-a)$ contains the interior of a fundamental domain
of $\Delta$,
$\sigma(b) \in \pi_{E^{\perp}}(\gamma-a)-\Delta$ for almost all
$b\in\pi_{E^{\perp}}(\gamma-a)$. If $\sigma'$ is strictly contracting on
$E\cap P^{\perp}$,
(\ref{self2}) may be satisfied for some finite power $\sigma = \sigma'^n$
provided all
$b\in\pi_{E^{\perp}}(\Lambda)\cap\pi_{E^{\perp}}(\gamma-a)$
are inner points of $\gamma-a\cap\{E+b\}$.

Therefore automorphisms of
the form
$\sigma=\sigma'^n$ with
$\sigma'=\prod_{k\in \Integer^*_h}p_k^{n_k}({\cal C})$,
$n_k\in \Integer$, which satisfy
\begin{equation}                              \label{Ba3}
\prod_{k\in \Integer^*_h}| p_k^{n_k}(\tau_h^{(m)}) | < 1
\end{equation}
for all $m\in \Integer^*_h\backslash\{1,-1\}$ are possible candidates for
inflations.
Any solution of the inequality (\ref{Ba3}) has to restrict
to an automorphism
of $E\cap P^{\perp}\cap\Lambda$, requiring therefore
$E\cap P^{\perp}\cap\Lambda=\{\vec{0}\}$.
(\ref{Ba3}) is equivalent to
\begin{equation}                             \label{Ba4}
\sum_{k\in \Integer^*_h\backslash\{1,-1\}} n_k ln|p_k(\tau^{(m)}_h)| < 0
\end{equation}
and may be certainly obtained, if the determinant of the matrix having
$ln|p_i(\tau^{(j)}_h)|$ as entries
($i,j\in \Integer^*_h\backslash\{1,-1\}$) does not vanish.
Note that again (\ref{Ba3}) does only refer to the Coxeter number,
not to the group itself.
It is not difficult to construct solutions of (\ref{Ba3}) by hand as
long as the number of elements in $\Integer^*_h$ is not to large.
By numerical inversion of
the above mentioned matrix a solution for $h=29$ was obtained.

\subsection{$\Integer^h$- versus $A_{h-1}$-Projections}

The set of classes of locally isomorphic tilings obtained from
$A_{h-1}$ on the one hand and from $\Integer^h$ on the other are
intimitely related.
The mapping $\alpha_i\mapsto e_i-e_{i+1}$ is an embedding of the
root lattice $\Lambda_{h-1}$ of $A_{h-1}$ into $\Integer^h$
in such a way that $\Integer^h \cap \Real\Lambda_{h-1} = \Lambda_{h-1}$.
It is the symmetry axis $P^{(0)}$ of the $h$-fold symmetry
which adds to $\Real\Lambda_{h-1}$ to yield the embedding
space $\Real^h$ of $\Integer^h$.
Moreover the $h$-fold symmetry of
$\Integer^h$ restricts to a Coxeter element $\omega'$ of $A_{h-1}$,
although it is not the canonical one. Hence
the plane along which the $\Integer^h$-strip is constructed
corresponds to $P^{(1)}_{\omega'}\subset \Real\Lambda_{h-1}$.
Let $\delta$ be the Weyl-reflection such that
\begin{equation}                                  \label{cox12}
\omega' = \delta\omega\delta^{-1} ,
\end{equation}
$\omega$ being the canonical Coxeter element.
By transforming the links as
${\cal L} \mapsto \delta^{-1}{\cal L}$
we may compare $A_{h-1}$-projections directly with $\Integer^h$-projections.
We have
\begin{equation}
 \pi^{\perp}(\Integer^h) =
 \pi^{\perp}(\Integer e_1) + \pi^{\perp}(\Lambda_{h-1}),
\end{equation}
and from $\pi^{\perp}(he_1) \subset
 \pi^{\perp}(\sum_{i=1}^h e_i) + \pi^{\perp}(\Lambda_{h-1})$ and
$\Integer\sum_{i=1}^he_i = P^{(0)}\cap \Integer^h$ one concludes
that
$\pi^{\perp}(n e_1) + \pi^{\perp}(\Lambda_{h-1})$ and
$\pi^{\perp}(m e_1) + \pi^{\perp}(\Lambda_{h-1})$ have for $n\neq m$
at least Euclidian distance $\sqrt{h}$. Therefore
\begin{equation}
\overline{ \pi^{\perp}(\Integer^h)} =
 \pi^{\perp}(\Integer e_1) + \overline{\pi^{\perp}(\Lambda_{h-1})},
\end{equation}
hence
\begin{equation}
\overline{ \pi^{\perp}(\Integer^h)}/ \pi^{\perp}(\Integer^h)
 = \overline{\pi^{\perp}(\Lambda_{h-1})}/\pi^{\perp}(\Lambda_{h-1}).
\end{equation}
This shows that the difference between the
quantum sets related to $\Integer^h$ on the one hand and
to $A_{h-1}$ on the other may have its origine only in the topoplogy
of the total space which is after all determined by the singular tilings.

As mentioned above the Coxeter element $\omega'$
is not the canonical one we discussed in (\ref{b3}).
In fact for $A$-Coxeter groups there are two natural choices for
Coxeter elements
(up to inversion).
The one here is $\omega' = s_1 s_2 \cdots s_{h-1}$
having numbered the simple roots, i.e. the vertices of the
Coxeter graph in rising order from one end to the other (which
is possible only for $A$-Coxeter groups).
For this reason the $A_{h-1}$-projections
look quite different from $\Integer^h$-projections.
In case of the Coxeter group $A_4$, there is an interesting connection
with Ammann bars as they are described in \cite{GrSh}
\cite{Lev} for Penrose tilings.
Ammann bars are a decoration of the prototiles out of which one may
construct
a Penrose tiling. In fact there is an alternative procedure
(actually the original one) to construct such a tiling,
namely by fitting prototiles together, where one has to obey certain
local matching conditions. Ammann bars may be understood as a
way to formulate these conditions.
If one keeps only the Ammann bar decoration of a complete tiling,
one is left with a so called Ammann-quasicrystal \cite{Lev},
which consists of five
$1$-dimensional tilings of the sort we obtained from our decomposition
of an $A_4$-tiling. However, according to a result
of John H. Conway, which is stated in \cite{GrSh},
these five $1$-dimensional tilings are not completely independent but
the whole Ammann-quasicrystal is determined, up to an uppermost
threefold degeneracy,
by just two of them. These two may be taken to be a $2$-dimensional $A_4$
tiling projected onto the canonical plane (\ref{b3}) and taking the
links to be the simple roots (an $A_4$-projection).
Figure 1 is meant to illustrate this.
It consists of the so-called "cartwheel tiling" \cite{GrSh} together
with its Ammann bar decoration appearing here as five aperiodic sequences of
parallels
with distances $1$ or $\tau$ (the golden ratio).
Figure 1 has a fivefold symmetry, but it is incomplete.
In the cartwheel tiling the so-called
Conway worms are not filled
out and correspondingly in each of the five
sequences of parallels
the line closest to the center is missing. If one now completes two of them
(there are two choices for each one) one is left
with only three (up to symmetry) instead of eight choices for completing
the others.
\bigskip

\noindent
For completeness
let us finally mention how candidates for inflations for $A_{h-1}$-tilings
may be used to define candidates for inflations for $\Integer^h$-tilings.

The role of the square of connectivity matrix is played by the matrix
${\cal M}+2id$, $\cal M$ having in the canonical basis entries
${\cal M}_{ij} = 1$ if $j=i-1\,mod\,h$ or
$j=i+1\,mod\,h$ and otherwise $0$.
E.g.\ for $h=5$
\begin{equation}
{\cal M}  =  \left( \begin{array}{ccccc}

              0 & 1 & 0 & 0 & 1   \\
              1 & 0 & 1 & 0 & 0   \\
              0 & 1 & 0 & 1 & 0   \\
              0 & 0 & 1 & 0 & 1   \\
              1 & 0 & 0 & 1 & 0

             \end{array} \right).
\end{equation}
Clearly this matrix commutes with the $h$-fold symmetry showing that it
leaves $P^{(m)}_{\omega'}$ invariant, $0\leq m \leq\frac{h}{2}$.
Note that ${\cal M}$ is the connectivity
matrix of the affine extension of $A_{h-1}$.
Its eigenvalues are
$2cos(\frac{2\pi m}{h})=4cos^2(\frac{\pi m}{h})-2$ for all $m$.
For this reason (and with the above identification $\alpha_i= e_i-e_{i+1}$)
\begin{equation}
({\cal M}+2id)|_{\Lambda_{h-1}} = \delta{\cal C}^2\delta^{-1}.
\end{equation}
Now let
$p = \prod_{k\in \Integer^*_h}p_k^{n_k}$, $n_k\in \Integer$,
be a solution of (\ref{Ba3})
and for which in addition holds
$p(2) = 1\,mod\,h$. Let $q$ be the polynomial defined by $q(x^2)=p(x)$
($p$ has to be even).
Then $q({\cal M}+2id)$ may be understood as an automorphism of
$\Integer^h/(P^{(0)}\cap \Integer^h)$ which is strictly
contracting on $E\cap P^{\perp}$, hence it is a canditate for an inflation
for $\Integer^h$-tilings.
This follows from $\pi(S(\gamma-\lambda)\cap \Integer^h) =
\pi(S(\gamma)\cap \Integer^h)$ for $\lambda\in P^{(0)}\cap \Integer^h$
and the fact that $2$
is the eigenvalue of $\cal M$ to $P^{(0)}$.
Now as $p_k(2)=k$
and $\Integer^*_h$ is finite, there is for $k\in\Integer_h^*$
always a power $n'$ such that $p_k^{n'}(2)=1\,mod\,h$.
Hence a given even solution of (\ref{Ba3})
leads always to a candidate for an inflation of $\Integer^h$.
As an example, for
$\Integer^7$ the polynomial
$p_2p_4 = x^4-2x^2$ yields an inflation which scaling factor 4.049.
It is given in figure~2.
\newpage

\section{Algebraic Characterization}

\subsection*{Characterization of Quantum Spaces by
$K_0$-Groups}

The notion of a noncommutative space (or a quantum space), as it is
investigated in A. Connes book \cite{Con}, arose from
the attempt to extend the duality between unital commutative
$C^*$-algebras
and compact Hausdorff spaces to noncommutative $C^*$-algebras.
This duality is expressed in the theorem of Gelfand-Neumark stating
that for a unital commutative $C^*$-algebra $A$
\begin{equation}                               \label{a5}
 A = C(spec(A)) ,
\end{equation}
i.e.\ $A$ coincides with the algebra of continuous complex functions over
its spectrum $spec(A)$, which is given by the set of all maximal
ideals of $A$ and which may be considered as a compact subspace
(in the weak*-topology) of $A'$, the dual
vector space of $A$. Moreover for
compact $X$, $spec(C(X)) = X$ as a topological space.
Hence for a complete extension of the above duality to
noncommutative algebras,
an appropriate notion
of their dual (generalizing the commutative case)
has to be set up as well as a construction which, for a given space,
yields an algebra such that the space appears as the dual of this algebra.

The notion of a spectrum of an algebra extending the commutative
case which is used in \cite{Con} is defined with the help of the
irreducible representations
(irreps) of the algebra:
\begin{equation}                                   \label{a3}
 \hat{A} = \{\mbox{irreps of}\, A \}\: \mbox{modulo unitary equivalence} .
\end{equation}
An alternative notion could be the primitive spectrum
$ \check{A} = \{ Ker \rho\, |\, \rho\, \mbox{irrep of A}\}$,
however this turns out to be to small.
In general $\check{A}\subset\hat{A}$ and both sets are equal
if and only if $A$ is of type I (or postlimial) \cite{Ped}.

Instead of discussing the general theory let us specialize to spaces of
the form that have been obtained in the first part of this article,
namely spaces $X/R$, where $X$ is a compact totally disconnected
Hausdorff space and
$R$ is an equivalence relation. How can one construct an algebra $A$,
such that $\hat{A} = X/R$?

An equivalence relation carries a groupoid structure.
For elements $(x,y)\in R \subset X\times X$
composition and inversion are defined
through the transitivity and the
reflexivity: $(x,y)$ and $(y',z)$ are composable iff $y=y'$ yielding
$(x,y)(y,z) = (x,z)$, and $(x,y)^{-1}=(y,x)$.
A. Connes proposes to take $A$ to be the groupoid $C^*$-algebra $C^*(R)$
defined by $R$.
To define it, one first considers the complex continuous functions over $R$
with compact support and
(truncated) convolution product and involution, given in case the orbits
of $R$ are discrete by
\begin{eqnarray}                             \label{a1}
f*g\,(x,y) & = & \sum_{z\sim x} f(x,z)g(z,y) , \\
f^*\,(x,y) & = & \overline{f(y,x)} .
\end{eqnarray}
This is a topological $*$-algebra. The norm of an element may be defined
to be the supremum of its operator-norms in all
bounded $*$-representations. Closure with respect to this norm yields
the $C^*$-algebra $C^*(R)$ \cite{Ren}. For more general situations,
where one has to put a measure $R$, see \cite{Ren}.
Let us look at two simple examples: \\
1) If $R = \{(x,x)|\,x\in X\}$, i.e. $X/R=X$, then the product
in (\ref{a1}) becomes the usual point multiplication of continuous
functions over $X$
and we recover the commutative case. \\
2) Let $X = \{x_i^a\}_{i,a}$ with $i\in\{1,\cdots,k\}$ and
$a\in\{1,\cdots,m_i\}$
be a finite set and $R$ be the equivalence relation
$x_i^a \sim x_i^{a'}$ for $a,a'\in\{1,\cdots,m_i\}$.
Then the product (\ref{a1}) coincides with matrix multiplication
in
\begin{equation}                                \label{a2}
C^*(R) \cong \bigoplus_{i=1}^{k} M_{m_i}(\Complex).
\end{equation}
$X/R$ has $k$ points, each one may be identified with an irrep of
$C^*(R)$, i.e.\ each one corresponding to a simple component of $C^*(R)$.

The second example already shows that $C^*(R)$ contains not only information
about the space $X/R$, but also about the largeness of the orbits
(equivalence classes) of $R$.
But as long as we are interested in the quotient space only,
and not its realization as a specific quotient,
only the invariants of $C^*(R)$ which are insensitive against a change of
such a realization characterize the quotient space,
among these being its $K_0$-group.
In fact $K_0$-groups are invariants of stabilized algebras only, and this
is reflected in example 2 above by the fact that
$C^*(R)$ and $M_n(C^*(R))\cong C^*(R)\otimes M_n(\Complex)$
have the same spectrum.
The stabilized algebra of a $C^*$-algebra $A$ is by definition
$M_{\infty}(A)\cong A\otimes M_{\infty}(\Complex)$,
$M_{\infty}(\Complex)$ being the algebra of matrices with all but finitely
many entries being non zero.
It may be obtained by taking the direct (or induced) algebraic limit
of the system $M_n(\Complex) \subset M_{n+1}(\Complex) \subset \cdots$
of algebras, the embeddings being nonunital.

However, if $X/R$ is a quantum space of tilings, $C^*(R)$ will
not be of type I, its primitive spectrum will reduce to one point and
its spectrum will be much larger than $X/R$, i.e. the
"non-commutative" analog of (\ref{a5}) fails to hold true.
Nevertheless $C^*(R)$ -
or better its $K_0$-group - will still characterize $X/R$. In a way
the functor $K_0$ singles out certain irreps of $C^*(R)$ which
allow to recover $X/R$ by restricting (\ref{a3}) to these irreps.

In general $K_0(A)$, the $K_0$-group of a unital $C^*$-algabra $A$,
is given
by the Grothendieck completion of the following abelian monoid $V(A)$:
$V(A)$ consists of equivalence classes of
projections in the stabilized algebra $M_{\infty}(A)$, namely $p\sim q$
whenever there are elements
$x,y\in M_{\infty}(A)$ such that $p=xy$ and $q=yx$.
The sum of projections $p+q$ is again a projection, if they are orthogonal,
i.e. $pq=0$. However, by using the stabilized algebra
$M_{\infty}(A)$ instead of $A$ itself, it is always
possible to find, for given classes $[p]$, $[q]$, orthogonal representatives
$\tilde{p}$, $\tilde{q}$ yielding a well defined addition
$[p]+[q]=[\tilde{p}+\tilde{q}]$ on $V(A)$.
$K_0(A)$ itself is given by formal differences in $V(A)$, i.e. its elements
are given by pairs $([p],[q])\in V(A)\times V(A)$ modulo the equivalence
relation $([p],[q])\sim ([p'],[q'])$ whenever there is a $[r]\in V(A)$
such that $[p]+[q']+[r] = [q]+[p']+[r]$.

Not only the $K_0$-group but
also $K_0^+(A) = \{([p],[0])|\,[p]\in V(A)\}$ characterizes the algebra,
as the latter elements
correspond to projections in $M_{\infty}(A)$.
To avoid cumbersome notation we shall write $[p]$ for $([p],[0])$.
In the cases which are
of interest for us, $K_0^+(A)$ defines a
positive cone, i.e.\ an additively closed subset such that
$K_0^+(A)-K_0^+(A)=K_0(A)$ and $K_0^+(A)\cap -K_0^+(A)=\vec{0}$. In other
words $K_0(A)$ becomes an ordered group.
The image of the unit in $K_0^+(A)$ is called the order unit.

We will also have to use the $K_1$-group of $A$. For its definition consider
$GL_n(A)$, the group of invertible elements of $M_n(A)$. $GL_n(A)$ may
be embedded into $GL_{n+1}(A)$ by a group homomorphism yielding
a directed system of groups the direct limit of which shall be denoted
by $GL(A)$. Let $GL(A)_0$ denote the connected component of the unit.
Then $K_1(A) = GL(A)/GL(A)_0$. $K_1(A)$ is an abelian group which is
trivial if $A$ is an $AF$-algebra.

For a reference to general operator $K$-theory see \cite{Bla}.
In the context of $AF$-algebras, which is discussed in more detail below,
the above abstract definition yields an intuitive picture
which may be encoded in a Bratteli diagram. According to a
theorem of Elliot \cite{Ell},
the $K_0$-group together with an order structure is
a complete invariant for the stabilized algebra of an $AF$-algebra, and
two $AF$-algebras having the same stabilized algebra are isomorphic, if
their order units coincide.

\subsection{Application to $ADE$-tilings}    \label{sub21}

The strip method suggested a realization of the quantum space of
tilings by
\begin{equation}                     \label{qu1}
\Psi = \Omega/\hbox{transl.},
\end{equation}
but here an orbit of the equivalence relation, i.e.\ an equivalence class
of a tiling, contains not all translates
of that tiling but only those which are integer linear combinations
of links. We denote the groupoid defined be the equivalence relation by
$R_{\Omega}$.
Taking, as at the end of subsection~\ref{ss1},
   ${\cal L}=\{\alpha_i\}_i$ application of
theorem~\ref{b4} leads to a decomposition, first of the total space,
$\Omega=\Omega_1\times \Omega_2$, and second of the equivalence classes,
$R_{\Omega} = R_{\Omega_1}\times R_{\Omega_2}$,
so that
\begin{equation}                    \label{psi}
\Psi = \Psi_1\times\Psi_2,
\end{equation}
where $\Psi_{\epsilon}$ is related to
$(\Lambda_{\epsilon},P_{\epsilon},{\cal L}_{\epsilon})$.
\bigskip

\noindent
Let us first consider one single factor $\Psi_{\epsilon}$. Such a situation
has been investigated by J. Bellissard \cite{B}.
Remember that the $1$-dimensional tilings appear
as twosided sequences of prototiles, which are in this case intervals of
the form
$(\pi(x),\pi(x+\alpha_i))$, $\alpha_i\in{\cal L}_{\epsilon}$,
$x\in \Lambda_{\epsilon}$.
Such a prototile shall be abbreviated by $i$
and referred to as a {\em letter}.
A natural topology on $\Omega_{\epsilon}$ is given by the following
subbase \cite{Q}:
Its sets are labelled by a letter $i$ and $k\in \Integer$:
\begin{equation}
U_{i}^{(k)} = \{{\cal T}\,|\hbox{ the $k$'th letter of ${\cal T}$ is $i$}\} .
\end{equation}
Here we say that the first letter is the one to the right of $\vec{0}$.
Since $U_i^{(k)}\cap U_j^{(k)} = \emptyset$ if $i\not= j$, and
$\Omega_{\epsilon}\backslash U_i^{(k)} = \bigcup_{j\not= i} U_j^{(k)}$,
these sets are open and closed so that $\Omega_{\epsilon}$ becomes a
totally disconnected space.
In fact this topology is inherited from the product topology on the set of
all twosided sequences of the above letters. That set is compact and
metrizable. A distance between
two such sequences $\{i_j\}_{j\in\Integer}$ and $\{i'_j\}_{j\in\Integer}$
may be defined by $exp(-sup\{r\,|\,\forall |j|\leq r:\,i_j=i'_j\})$.
As a consequence of the definition of the $LI$-class $\Omega_{\epsilon}$
contains all its limit points and hence is compact.
Now any orbit of the groupoid $R_{\Omega_{\epsilon}}$ may be
identified with $\Integer$, since translation here amounts to shifting
by a certain number of letters.
A shift by one letter to the right is a topologically transitive
homeomorphism on $\Omega_{\epsilon}$. We denote this shift action by
$\hat{\varphi}_{\epsilon}$ and
identify $R_{\Omega_{\epsilon}}$ with $\Omega_{\epsilon}\times \Integer$
through $({\cal T},\hat{\varphi}_{\epsilon}^{-k}({\cal T}))
\cong ({\cal T},k)$.
Then $\Omega_{\epsilon}$ is the closure of one single
$\Integer$-orbit.

The groupoid $C^*$-algebra defined by $\Omega_{\epsilon}\times \Integer$
is a crossed product of $C(\Omega_{\epsilon})$ with $\Integer$,
namely (\ref{a1}) becomes
\begin{equation}                             \label{a10}
f*g\:({\cal T},k)  = \sum_{m\in\Integer} f({\cal T},m)\,
g(\hat{\varphi}_{\epsilon}^{-m}({\cal T}),k-m) ,
\end{equation}
so that the functions $\hat{f}:\Integer\longrightarrow C(\Omega_{\epsilon})$
given by $\hat{f}(k)({\cal T}) = f({\cal T},k)$ are elements of
\begin{equation}                       \label{ps1}
C(\Omega_{\epsilon}) \times_{\varphi_{\epsilon}} \Integer
\end{equation}
with involution now given by
$\hat{f}^*\,(k) = \overline{\varphi_{\epsilon}^k(\hat{f}(-k))}$.
Here the action $\varphi_{\epsilon}$ on $C(\Omega_{(\epsilon)})$
is the pull back of $\hat{\varphi}_{\epsilon}$, i.e.\
$\varphi_{\epsilon}(\hat{f}) = \hat{f}\circ\hat{\varphi}_{\epsilon}^{-1}$.
Bellissard's work contains also an approach to this algebra which is
physically motivated. The single tiling is then understood as a quasicrystal
and the above algebra is related to the algebra of observables.
The $K$-groups of
$C(\Omega_{\epsilon}) \times_{\varphi_{\epsilon}} \Integer$ have been
computed e.g.\ in \cite{BBG} with the help of
the Pimsner Voiculescu exact sequence \cite{PV}.
This sequence exists for any $C^*$-algebra $A$ and given $\alpha\in Aut(A)$
leading to an action of $\Integer$ on $A$ (by iteration). It is the cyclic
exact sequence
\begin{equation}                       \label{pv1}
\begin{array}{ccccc}
K_0(A) &\stackrel{id - \alpha_*}{\longrightarrow} &
K_0(A) &\stackrel{i_*}{\longrightarrow} &
K_0(A\times_{\alpha}\Integer)    \\
\uparrow & & & & \downarrow       \\
K_1(A\times_{\alpha}\Integer)  &\stackrel{i_*}{\longleftarrow} &
K_1(A) &\stackrel{id - \alpha_*}{\longleftarrow} &
K_1(A)
\end{array}
\end{equation}
$i$ denoting the canonical injection from $A$ into
$A\times_{\alpha}\Integer$. (Any algebra homomorphism $\phi$ induces a unique
group
homomorphism on the corresponding $K$-groups commonly denoted by $\phi_*$.)
Applied to $A=C(\Omega_{\epsilon})$
and $\alpha = \varphi_{\epsilon}$ the following result is obtained:
As $\Omega_{\epsilon}$ is a totally disconnected compact and
metrizable space,
$C(\Omega_{\epsilon})$ is an AF-algebra \cite{HPS},
$K_0(C(\Omega_{\epsilon}))\cong
C(\Omega_{\epsilon},\Integer)$,
$K_0^+(C(\Omega_{\epsilon}))\cong
\{f\in C(\Omega_{\epsilon},\Integer)|\,f\geq0\}$, and
$K_1(C(\Omega_{\epsilon}) = \{0\}$,
$C(\Omega_{\epsilon},\Integer)$ denoting the continous functions
over $\Omega_{\epsilon}$ with values in $\Integer$. Under the above
identification $\varphi_{\epsilon\,*} f=f\circ\hat{\varphi}_{\epsilon}^{-1}$.
Therefore (\ref{pv1}) leads to
\begin{eqnarray}                                  \label{pv30}
K_0(C(\Omega_{\epsilon})\times_{\varphi_{\epsilon}}\Integer)&\cong&
C(\Omega_{\epsilon},\Integer)/E_{\varphi_{\epsilon}} \\ \label{pv40}
K_1(C(\Omega_{\epsilon})\times_{\varphi_{\epsilon}}\Integer)&\cong&
\Integer ,
\end{eqnarray}
where $E_{\varphi_{\epsilon}} = (id-\varphi_{{\epsilon}\,*})
(C(\Omega_{\epsilon},\Integer))$ and
$ker(id-\varphi_{\epsilon\,*})=\Integer$ since $\hat{\varphi}_{\epsilon}$
acts transitively on $\Omega_{\epsilon}$.
This consideration does not lead, at least not directly, to the determination
of the positive cone, but it is shown e.g.\ in \cite{HPS} that
$K_0^+(C(\Omega_{\epsilon})\times_{\varphi_{\epsilon}}\Integer)$
is given by the image of
$\{f\in C(\Omega_{\epsilon},\Integer)|\,f\geq0\}$ in
$C(\Omega_{\epsilon},\Integer)/E_{\varphi_{\epsilon}}$.
\smallskip

\noindent
Before we proceed to the $2$-dimensional case, let us briefly illustrate
the gap labelling theorem \cite{B,BBG}:
Let $h:R_{\Omega_{\epsilon}}\longrightarrow \Complex$ be a selfadjoint
element of
$C^*(R_{\Omega_{\epsilon}})\cong
C(\Omega_{\epsilon}) \times_{\varphi_{\epsilon}} \Integer$ and
consider for fixed
${\cal T}\in\Omega_{\epsilon}$ the representation $\pi_{\cal T}$ on
$\ell^2(\Integer)$
\begin{equation}
\pi_{\cal T}(f)\psi(k) = \sum_{m\in\Integer}
f(\hat{\varphi}_{\epsilon}^{-k}({\cal T}),m-k)\,\psi(m) .
\end{equation}
For reasonable chosen $h$, $H=\pi_{\cal T}(h)$ may be considered as a
$1$-dimensional Hamiltonian which acts on wavefunctions over
the tiling ${\cal T}$.
Let $\chi_{H\leq E}$ denote the projection onto
the subspace spanned by the eigenfunctions of $H$ of
energy less or equal then $E$.
For most of the energy values it does not belong to the $C^*$-algebra above,
but it is argued that,
if $E$ is the lower value of a spectral gap of the Hamiltonian,
indeed $\chi_{H\leq E}\in \pi_{\cal T}(C^*(R_{\Omega_{\epsilon}}))$.
Classes of projectors are in a sense stable against small perturbations,
namely $\|p-q\|<1$ implies $[p]=[q]$.
This explains the stability of the gap labelling,
because the spectral gaps having
lower energy $E$ may be labelled
by $tr(\chi_{H\leq E})=tr_*([\chi_{H\leq E}])$.
Hereby $tr_*:K_0(C^*(R_{\Omega_{\epsilon}}))\longrightarrow \Real$ is the
tracial state induced by the normalized trace on
$C^*(R_{\Omega_{\epsilon}})$. The latter restricts to a
$\hat{\varphi}_{\epsilon}$-invariant trace on $C(\Omega_{\epsilon})$,
i.e.\ it is determined by an invariant measure on
$\Omega_{\epsilon}$. Therefore the elements of
$tr_*(K_0(C^*(R_{\Omega_{\epsilon}})))\cap [0,1]$
are given by integer linear combinations of
 relative frequencies
of patterns (words) in the tiling. In certain cases (see below) these values
already determine the ordered $K_0$-group completely.
One expects that,
if the potential term in $H$ depends
upon the structure (e.g.\ the prototiles) of the tiling, most of the values
of $tr_*$ between $0$ and $1$ do actually correspond to labels of spectral
gaps of $H$.
\bigskip

\noindent
That the quantum space in the $2$-dimensional case is a cartesian
product implies that relative frequencies of $2$-patterns are
just given by products of relative frequencies of $1$-patterns.
However, a similar result as in the purely $1$-dimensional case, namely
that the relative frequences of
patterns determine the  gap labelling or even the ordered $K_0$-group,
is a priori not clear.
The groupoid being $(\Omega_1\times \Integer)\times(\Omega_2\times\Integer)$
its $C^*$-algebra is now a crossed product with
$\Integer\oplus\Integer$,
\begin{equation}                                  \label{ps2}
C(\Omega) \times_{\varphi_1 \times \varphi_2}
\Integer\oplus\Integer,
\end{equation}
where $\varphi_1$ and $\varphi_2$
are the pull back action of the first resp.\ second $\Integer$.
Here $\hat{\varphi}_{\epsilon}$ acts trivially on
$\Omega_{\epsilon+1}$. We may as well express (\ref{ps2}) as
\begin{equation}                                  \label{ps3}
(C(\Omega) \times_{\varphi_1}\Integer)
\times_{\varphi_2}\Integer,
\end{equation}
$\varphi_2(b)(m)=\varphi_2(b(m))$ for
$b\in C(\Omega) \times_{\varphi_1}\Integer$
and therefore apply the Pimsner Voiculescu exact sequence
twice to obtain the $K$-groups.
\begin{thm} \label{thm5}
Let $\Omega_{\epsilon}$, $\epsilon = 1,2$, be a totally disconnected
compact metrizable spaces and $\varphi_{\epsilon}$ topologically transitive
 homeomorphisms on $\Omega_{\epsilon}$.
Then, with $\Omega = \Omega_1 \times \Omega_2$,
\begin{equation}                      \label{pv6}
K_0((C(\Omega)\times_{\varphi_1}\Integer)
     \times_{\varphi_2}\Integer)
\cong C(\Omega_1,\Integer)/E_{\varphi_1} \otimes
C(\Omega_2,\Integer)/E_{\varphi_2} \,\oplus\, \Integer,
\end{equation}
and, if $C(\Omega_1,\Integer)/E_{\varphi_1}$ is free,
\begin{equation}                      \label{pv8}
K_1((C(\Omega)\times_{\varphi_1}\Integer)
     \times_{\varphi_2}\Integer)
\cong C(\Omega_1,\Integer)/E_{\varphi_1} \oplus
C(\Omega_2,\Integer)/E_{\varphi_2}.
\end{equation}
\end{thm}
{\em Proof:} We first apply the Pimsner Voiculescu exact sequence
to $A=C(\Omega)$
and $\alpha = \varphi_1$. As $\Omega$ is totally
disconnected compact and metrizable,
one obtains, analogous to the purely $1$-dimensional
case, $K_0(C(\Omega))\cong
C(\Omega,\Integer)$ and
$K_1(C(\Omega)) = \{0\}$.
Hence (\ref{pv1}) yields an exact sequence
\begin{eqnarray}                       \label{pv2}
0\longrightarrow
K_1(C(\Omega\times_{\varphi_1}\Integer))
 \longrightarrow
C(\Omega,\Integer)
 \stackrel{id - \varphi_{1\,*}}{\longrightarrow}
C(\Omega,\Integer)
\stackrel{i_*}{\longrightarrow}
K_0(C(\Omega)\times_{\varphi_1}\Integer)
\longrightarrow  0                       \nonumber
\end{eqnarray}
and under the above identifications $\varphi_{\epsilon\,*}$ becomes the
pull
back of $\hat{\varphi}_{\epsilon}$.
Consequently
\begin{eqnarray}                                  \label{pv3}
K_0(C(\Omega)\times_{\varphi_1}\Integer)&\cong&
C(\Omega,\Integer)/ im (id - \varphi_{1\,*})\cong
C(\Omega_1,\Integer)/E_{\varphi_1} \otimes C(\Omega_2,\Integer) \\
\label{pv4} K_1(C(\Omega)\times_{\varphi_1}\Integer)&\cong&
ker (id - \varphi_{1\,*}) = C(\Omega_2,\Integer) .
\end{eqnarray}
Hereby use has been made of
$C(\Omega)\cong C(\Omega_1) \otimes C(\Omega_2)$
($C(\Omega)$ is an AF-algebra).
Moreover $\hat{\varphi}_1$ being topologically transitive implies
that every $\hat{\varphi}_1$-invariant function on $\Omega_1$ is
constant which leads to
(\ref{pv4}). Note that ($i=0,1$)
$K_i(C(\Omega)\times_{\varphi_1}\Integer)\cong
K_i(C(\Omega_1)\times_{\varphi_1}\Integer) \otimes C(\Omega_2,\Integer)$.
Application of the Pimsner Voiculescu sequence to
$A = C(\Omega) \times_{\varphi_1}\Integer$
and $\alpha = \varphi_2$ now gives
$$
\begin{array}{ccccc}
C(\Omega_1,\Integer)/E_{\varphi_1} \otimes C(\Omega_2,\Integer)
&\stackrel{id - \varphi_{2\,*}}{\longrightarrow} &
C(\Omega_1,\Integer)/E_{\varphi_1} \otimes C(\Omega_2,\Integer)
&\stackrel{i_*}{\longrightarrow} &
K_0(A \times_{\varphi_2}\Integer)    \\
\uparrow & & & & \downarrow       \\
K_1(A \times_{\varphi_2}\Integer)  &\stackrel{i_*}{\longleftarrow} &
C(\Omega_2,\Integer) &\stackrel{id - \varphi_{2\,*}}{\longleftarrow} &
C(\Omega_2,\Integer)
\end{array}
$$
The kernel of $id - \varphi_{2\,*}$ as an endomorphism on
$C(\Omega_2,\Integer)$ is $\Integer$, a free abelian group,
so that the exact sequence of abelian groups
$$
C(\Omega_1,\Integer)/E_{\varphi_1} \otimes C(\Omega_2,\Integer)
\stackrel{id - \varphi_{2\,*}}{\longrightarrow}
C(\Omega_1,\Integer)/E_{\varphi_1} \otimes C(\Omega_2,\Integer)
\stackrel{i_*}{\longrightarrow}
K_0(A \times_{\varphi_2}\Integer)
\longrightarrow \Integer \longrightarrow 0
$$
splits thus yielding (\ref{pv6}).
By the same reasoning one obtains
from
$$
C(\Omega_2,\Integer)
\stackrel{id - \varphi_{2\,*}}{\longrightarrow}
C(\Omega_2,\Integer)
\stackrel{i_*}{\longrightarrow}
K_1(A \times_{\varphi_2}\Integer)
\longrightarrow C(\Omega_1,\Integer)/E_{\varphi_1}\longrightarrow 0
$$
the result (\ref{pv8}), if $C(\Omega_1,\Integer)/E_{\varphi_1}$
is free (to guarantee the split property).\hfill $\Box$
\bigskip

\noindent
It is not known to us what the positive cone
$K_0^+((C(\Omega)\times_{\varphi_1}\Integer)
     \times_{\varphi_2}\Integer)$ is.
However, for application to the $2$ dimensional gap labelling,
refering to selfadjoint elements
$h:R_{\Omega}\longrightarrow \Complex$ of
$C^*(R_{\Omega})\cong (C(\Omega)\times_{\varphi_1}\Integer)
    \times_{\varphi_2}\Integer)$
which play the role of Hamiltonians in representations
on $\ell^2(\Integer^2)$, only
the values of the tracial states on the $K_0$-group need to be known.
These may be obtained by application of theorem~3 of \cite{Pi}:
It states that, given a trace $tr$ on $A\times_{\alpha}\Integer$,
the map
$\underline{\Delta}^{\alpha}_{tr}:ker(id-\alpha_*)\subset K_1(A)
\longrightarrow\Real/tr_*(K_0(A))$ defined by
\begin{equation}
\underline{\Delta}^{\alpha}_{tr}([u]) =
\frac{1}{2\pi i}\int_0^1 tr(\dot{\xi}(t)\xi^{-1}(t))\ dt\:/\:
tr_*(K_0(A)) ,
\end{equation}
$\xi:[0,1]\longrightarrow GL(A)$ being a piecewise contineous path from
$1$ to $u\,\alpha(u^{-1})$, is a well defined group homomorphism, and
moreover that the sequence
\begin{equation}
0\longrightarrow tr_*(K_0(A))\longrightarrow
tr_*(K_0(A \times_{\alpha}\Integer))
\stackrel{q}{\longrightarrow}
\underline{\Delta}^{\alpha}_{tr}(ker(id-\alpha_*))\longrightarrow 0 ,
\end{equation}
$q:\Real\longrightarrow\Real/tr_*(K_0(A))$ being the canonical
projection, is exact. Hereby the restriction of $tr$ to $A$ is also
denoted by $tr$. This enables us to show
\begin{thm}
Let $\Omega_{\epsilon}$ and $\varphi_{\epsilon}$ be as in the
theorem~\ref{thm5} and $tr$ be a trace on
$C(\Omega)\times_{\varphi_1}\Integer$. Then
\begin{equation}
tr_*(K_0((C(\Omega)\times_{\varphi_1}\Integer)\times_{\varphi_2}\Integer)) =
tr_*(C(\Omega,\Integer)).
\end{equation}
\end{thm}
{\em Proof:} We apply the abovementioned theorem to
$A = C(\Omega)\times_{\varphi_1}\Integer$ and $\alpha=\varphi_2$.
$ker(id-\varphi_{2\,*})\subset K_1(A)=C(\Omega_2,\Integer)$
was already determined to be
$\Integer$ and we need to identify a nontrivial element.
Consider the
constant function $u:\Integer\longrightarrow
C(\Omega)$ given by $u(n)=\delta_{1n}$ (Kronecker's $\delta$).
It is the unitary in $C(\Omega)\times_{\varphi_1}\Integer$
conjugation by which yields the action of $\varphi_1$:
\begin{equation}
u*f*u^{-1} = \varphi_1(f) ,
\end{equation}
$f\in C(\Omega)\times_{\varphi_1}\Integer$.
Moreover $u\notin GL(A)_0$, because otherwise $\varphi_1$
would be homotopic to
the identity. Since $\varphi_2(u)=u$,
$[u]$ is indeed a nontrivial element of
$ker(id-\varphi_{2\,*})\subset K_1(A)$.
Now $u\,\varphi_2(u^{-1}) = 1$ implies
$\underline{\Delta}^{\varphi_2}_{tr}([u]) = 0$ so that
\begin{equation}                          \label{bb1}
tr_*(K_0((C(\Omega)\times_{\varphi_1}\Integer)\times_{\varphi_2}
\Integer)) = tr_*(K_0(C(\Omega)\times_{\varphi_1}\Integer)) .
\end{equation}
The restriction of $tr$ to $C(\Omega)\times_{\varphi_1}\Integer$
has to be invariant under $\varphi_2$.
A second application of the above sequence
with $A=C(\Omega)$, $\alpha=\varphi_1$ directly implies
$tr_*(K_0(C(\Omega)\times_{\varphi_1}\Integer)) =
 tr_*(K_0(C(\Omega))$, the restriction of $tr$ to
$C(\Omega)$ now being $\varphi_1\times\varphi_2$-invariant.\hfill $\Box$
\bigskip

\noindent
Therefore (\ref{bb1}) is
determined by a $\varphi_1 \times \varphi_2$-invariant probability measure
on $\Omega$.
Moreover $tr_*(K_0(C(\Omega))=tr_*(C(\Omega_1,\Integer))\,
tr_*(C(\Omega_2,\Integer))$,
i.e.\ the invariant probability measure on $\Omega$ is the product
of the invariant measures on the single components (results of \cite{Q}
for the $1$-dimensional case show that it is unique).
Like in the $1$-dimensional case the values of the tracial state are given
by integer linear combinations of relative frequencies of $2$-patterns.

In \cite{BBG}, in case $\Omega_{\epsilon}$ is given by
substitution sequences, such an invariant measure is determined
using special technics related to these substitutions.
In the next section a different approach to their determination is proposed,
which has the benefit of making the ordered $K_0$-group itself more
transparent.

\subsection{Alternative description of the quantum space}

In his treatment of the class of locally isomorphic Penrose tilings,
i.e.\ the quantum space related to $(\Integer^5,P^{(1)})$, A. Connes
uses an alternative description which leads to another
realization of that space as a quotient, see also \cite{GrSh} for details.
It is based on local inflation/deflation rules.
By use of these rules to any Penrose
tiling together with a chosen (starting-) point
a sequence $\{f_i\}_{i\geq 0}$ may be assigned,
which has values in $\{0,1\}$ and satisfies
the constraint $f_i=1 \Rightarrow f_{i+1}=0$.
Translation of that tiling, i.e.\ a change of the starting point, amounts
to a change of the sequence yet only up to finitely many elements.
Therefore the quantum space of Penrose tilings is as well given by
the set of all $0,1$-sequences with the above constraint modulo the
equivalence relation which identifies sequences, if they differ only up to
finitely many elements. The above sequences may be visionalized as paths
over the graph $A_4/\Integer_2$,
or paths on the Bratteli diagram the
inclusion graph of which is $A_4$.
The groupoid-$C^*$-algebra is in this case the $AF$-algebra which is
defined by the Bratteli diagram. Ordered $K_0$-groups
of this type of algebras are well known and relatively easy to compute
(see below).

Comparing the approach of Connes with the statements made above,
one is led to the question:
\begin{itemize}
\item
Does this alternative way of describing
the quantum space of tilings carry over to the general case?
\end{itemize}
A step towards an answer to that question is given in \cite{HPS}, however
only in the one dimensional context.
The $\Omega_{\epsilon}$ together with the homeomorphism $\varphi_{\epsilon}$
furnishes a minimal topological dynamical system, and
$C(\Omega_{\epsilon}) \times_{\varphi_{\epsilon}} \Integer$
is the algebra
naturally assigned to it. Herman et al.\ found a way to formulate
such a dynamical system on a Bratteli diagram, the $\Integer$-action
being encoded in an order on that diagram,
hereby specifying one point in $\Omega_{\epsilon}$
(more precisely one has to consider equivalence classes of pointed
topological essential minimal dynamical systems on totally disconnected
compact spaces
and correspondingly about equivalence classes of
essentially simple ordered Bratteli diagrams).
$C(\Omega_{\epsilon}) \times_{\varphi_{\epsilon}} \Integer$
may roughly be understood as being generated by the $AF$-algebra defined
by the
Bratteli diagram and one additional element. Moreover it is shown that the
ordered $K_0$-group of the crossed product equals the ordered $K_0$-group
of that $AF$-algebra.

However, the construction of an ordered Bratteli diagram corresponding to a
pointed topological dynamical system
as it is described by Herman et al.\ is quite involved.
On the other hand, given an inflation $\sigma$ for some representative
$T\in\Psi_{\epsilon}$, it
naturally defines a Bratteli diagram,
and moreover it is possible - very similar to the Penrose-case -
to map at least the whole class of $\cal T$, i.e.\ the $\Integer$-orbit
 through $\cal T$
onto that diagram.
This encourages us to conjecture that the
the ordered $K_0$-groups of
$C(\Omega_{\epsilon}) \times_{\varphi_{\epsilon}} \Integer$
and the
$AF$-algebra being defined now by means of an inflation are equal.
\bigskip

\noindent
Let us first clarify the notions.
The inflation is a positive map and may therefore,
like the matrix $\cal C$,
be interpreted as a connectivity
matrix of some graph $\Sigma$ (which will in general not be a Coxeter
graph).
Again the vertices
of $\Sigma$ may be identified with the simple roots,
the $i$'th and the $j$'th vertex yet being now
joint by $\sigma_{ji}$ oriented edges, which are to be distinguished.
By saying that an edge $f$ is oriented, one means that it has a source
and a range denoted by the
maps $s$ and $r$, namely in the above example $s(f)=i$ and $r(f)=j$.
A path of length
$k$ over $\Sigma$ is given by a sequence of $k$ oriented edges
$f_1^{l_1},f_2^{l_2},\cdots,f_k^{l_k}$ such
that $r(f_n^{l_n})=s(f_{n+1}^{l_{n+1}})$.
We use extra labels $l_k$ in order to distuinguish between
the $\sigma_{ji}$ edges having source $i$ and range $j$.
The set of all infinite paths over $\Sigma$ shall be
denoted by ${\cal P}(\Sigma)$.
We did require that the inflation
decomposes into two parts $\sigma_1$, $\sigma_2$ (its
restrictions to $\Lambda_1$, $\Lambda_2$),
which then have to be irreducible. Hence $\Sigma$ has two connected
components  $\Sigma_1$, $\Sigma_2$. As long as we discuss
the $1$-dimensional case we consider one component only.
To better visionalize an infinite path over a connected
component $\Sigma_{\epsilon}$ it may be unfolded, the set of all infinite
paths
yielding a Bratteli diagram.

Bratteli invented these diagrams in order to describe $AF$-algebras
(approximately finite algebras). An $AF$-algebra is the $C^*$-hull of the
direct limit of a directed system
\begin{equation}                          \label{k2}
A_0 \stackrel{h_0}{\longrightarrow} A_1
\stackrel{h_1}{\longrightarrow} \cdots
\end{equation}
of finite dimensional $C^*$-algebras $A_n$ and
$*$-homomorphisms $h_n$; we denote it by $A_{\infty}$.
For our purposes it suffices
to consider unital embeddings only, i.e.\
$h_n$ embeds $A_n$ into $A_{n+1}$ preserving the unit.
 A Bratteli diagram may be understood
as a $K$-theoretic
description of that system together with an order unit in terms of an
infinite weighted graph. It determines the $AF$-algebra up to
$*$-isomorphisms.
The $K_0$-group $K_0(A_{\infty})$ of $A_{\infty}$ is
the direct limit of the directed system
of the $K_0$-groups of $A_n$, the positive homomorphisms $h_{n\,*}$
being induced by the above embeddings $h_n$.
Every finite dimensional $C^*$-algebra is semisimple, i.e.\ its elements
may be identified with block-diagonal matrices
\begin{equation}
A_n \cong \bigoplus_{k=1}^{k_n} M_{m_k}(\Complex) ,
\end{equation}
$M_{m_k}(\Complex)$ denoting the blocks consisting of
$m_k\times m_k$-matrices. Then $K_0(A_n) = \Integer^{k_n}$ the generators
standing for the distinct blocks or
minimal central projections of $A_n$, and
$h_{n\,*}\in Hom(\Integer^{k_n},\Integer^{k_{n+1}})$ is the
$k_{n+1}\times k_n$ matrix with entries in $\Integer^+$ its $ji$-coefficient
telling how often the block $M_{m_i}$ of $A_n$
is embedded into the block
$M_{m_j}$ of $A_{n+1}$ by $h_n$.
This is graphically encoded in a Bratteli diagram as follows:
The set of its vertices $V$ is grouped into floors $V_n$,
$V=\cup_{n\geq-1} V_n$, each
vertex of $V_n$ representing a generator
of $K_0(A_n)$. Its (oriented) edges are given through the homomorphisms
$h_{n\,*}$:
The $i$'th vertex of $V_n$ is joint
with the $j$'th vertex of $V_{n+1}$ by $(h_{n\,*})_{ji}$ edges all having
their source at that $i$'th vertex of $V_n$ and
their range at that $j$'th vertex of $V_{n+1}$.
In the context of unital embeddings all information about
the isomorphism class of the directed system is encoded in the graph except
of the dimensions of the blocks of the first algebra $A_0$. These may
be encoded by weights put on the vertices of $V_0$.
It is convenient to enlarge the directed system (\ref{k2}) by adding
$A_{-1}=\Complex\subset A_0$ to the left,
for then it is understood
that the dimension of the block of the first algebra (now being $A_{-1}$)
is $1$. In terms of the Bratteli diagrams one adds a floor $V_{-1}$
having one vertex representing $K_0(\Complex)=\Integer$ and $m_i$
edges from that vertex to the $i$'th vertex of $V_0$,
if the $i$'th block of $A_0$ is of size $m_i\times m_i$.
The weight at the vertex of $V_{-1}$
may then be left away, since it is always $1$.

The set of paths over $\Sigma_{\epsilon}$ yields such a Bratteli diagram:
$V_n=\Sigma^{(0)}_{\epsilon}$, for $n\geq 0$,
and the edges connecting the $i$'th
vertex of $V_n$ with the $j$'th of $V_{n+1}$ are given by the edges of
$\Sigma^{(1)}_{\epsilon}$ having their source resp. range at $i$ resp. $j$.
Adding the floor $V_{-1}$ amounts to indicating
the starting vertex of a single path over $\Sigma_{\epsilon}$, namely
by the edge that has
source at the single vertex of $V_{-1}$ and range equal the starting
vertex.
The directed system given now by the set of paths over $\Sigma_{\epsilon}$
is obtained by putting
for all $n\geq 0$
\begin{eqnarray}
 K_0(A_n) & = & \Lambda_{\epsilon} = \Integer^r ,\\
  h_{n\,*} & = & \sigma_{\epsilon} ,
\end{eqnarray}
$K_0(A_{-1}) = \Integer$ and $h_{-1\,*} = (1 1\cdots 1)^T$,
an $r\times 1$-matrix, $r = dim(\Lambda_{\epsilon})$.
Therefore
\begin{equation}                        \label{k1}
K_0(A_{\infty}) \cong \Lambda_{\epsilon} ,
\end{equation}
following from $\sigma_{\epsilon}$ being
an automorphism. Note that this group is free.
In terms of algebras, this means that $A_{\infty}$ is (the $C^*$-closure of)
the path algebra over the graph
$\Sigma_{\epsilon}$; we also denote it by $AF(\Sigma_{\epsilon})$.
If $\sigma_{\epsilon}$ is symmetric, it is approximated
by the tower
of the inclusion $\Complex^r=A_0\subset A_1$ with inclusion matrix
$\sigma_{\epsilon}$ \cite{GHJ}.
The order relation on the $K_0$-group is calculated as follows:
We may use the elements of $K_0(A_0)$ as representatives for elements
of the direct limit so that the positive cone is given by
\begin{equation}
K_0^+(A_{\infty}) = \bigcup_{n\geq 0}\sigma_{\epsilon}^{-n}(K_0^+(A_n)) .
\end{equation}
Hence $x\in K_0^+(A_{\infty})$ iff there is an $n\geq 0$ such that
$\sigma_{\epsilon}^n(x) \in K_0^+(A_0) =
(\Integer^+)^r$. Clearly if the
latter holds for $n$ it holds as well for $n+1$ as $\sigma_{\epsilon}$
is positive. For that reason it is the projection of $x$ along the
eigenspace to largest eigenvalue
of the inflation which is crucial, namely
\begin{equation}                                      \label{b10}
K_0^+(AF(\Sigma_{\epsilon})) = \{x\in \Lambda_{\epsilon}|
(\vec{\nu}_{\epsilon},x) > 0\}\cup\{0\} ,
\end{equation}
where $\vec{\nu}_{\epsilon}$ is the Perron Frobenius vector of
the inflation.
In particular, there is only one state on the ordered $K_0$-group
\cite{Eff},
which therefore has to be
$tr_*$, the homomorphism which is induced by the unique normalized
trace $tr: A_{\infty}\longrightarrow \Complex$. It is,
up to normalization, given by the orthogonal projection onto $P_{\epsilon}$.
If $P^{\perp}\cap\Lambda_{\epsilon}=\vec{0}$, $K_0(A_{\infty})$ is totally
ordered and any element of it is uniquely determined by its value under
$tr_*$.
As we already mentioned, the $AF$-algebra is determined by its
ordered $K_0$-group together with an order unit. The latter is necessary
for the definition of states (which involves a normalization)
but it is not invariant under a change
of a realization of $\Psi$.
Note that not only the ordered $K_0$-group is independent of a choice
for the inflation, as it should be, but even the above described
$AF$-algebra as
$(1,1,\cdots,1)\in\Integer^r$ may in all cases be taken to be
the order unit.
\bigskip

\noindent
Having chosen a
single tiling ${\cal T}={\cal T}_{\gamma_{\epsilon}-a_{\epsilon}}$
which allows for an
inflation $\sigma_{\epsilon}$ we can map the orbit through ${\cal T}$ onto
the Bratteli diagram, which has been described above.
As ${\cal T}$
appears as a twosided sequence of letters, its selfsimilarity
may be interpreted by saying that its letters are grouped into words,
each one corresponding to the
deflation (as the inverse process of inflation) of some letter of the
twosided sequence ${\cal T}_{\sigma_{\epsilon}
(\gamma_{\epsilon}-a_{\epsilon})}$.
More precisely consider, like in the proof of proposition~\ref{prop1},
points
$x,x'\in S_{\sigma_{\epsilon}(\gamma_{\epsilon}-a_{\epsilon})}
\cap\Lambda_{\epsilon}$
such that
$x'-x=\sigma_{\epsilon}(\alpha_i)=\alpha_{i_1}+\cdots+\alpha_{i_k}$
for some $\alpha_i\in {\cal L}_{\epsilon}$.
Through (\ref{seq3}) the order of the letters in the
word $i_1\cdots i_k$
is determined, and $(\sigma_{\epsilon})_{li}$ equals the number of times
the letter $l$ appears in that word.
Since the word arose by the process of deflation from $i$, we call
it an $i$-deflation. Strictly speaking the terminus letter-deflation
is a little bit unprecise, as the order of the letters in the word
described above does
not only depend on the letter but also on the point $\pi(x)$.
In case this order is independent of that point, the above
letter-deflation
furnishes what is called a
substitution, the tilings being substitution sequences,
see \cite{Q,BBG}.
In fact most of what follows may be applied directly to substitutions.

To any letter of ${\cal T}$ step by step a path over $\Sigma_{\epsilon}$
shall be assigned as follows:\\
{\em Step:}
Let $i_0$ be the letter to start with. It belongs to a unique
letter-deflation say of the letter $i_1$.
If the $i_0$ we have picked out is the $l_1$'s letter $i_0$ in that
$i_1$-deflation,
then we encode this by the $l_1$'th edge
 $f_1^{(l_1)}$
having $s(f_1^{(l_1)})=i_0$ and
$r(f_1^{(l_1)})=i_1$.

This step may be repeated, if one first uses (\ref{inf2}) to identify
${\cal T}_{\sigma_{\epsilon}(\gamma_{\epsilon}-a_{\epsilon})}$ with
${\cal T}_{\gamma_{\epsilon}-a_{\epsilon}}$ hereby
keeping track of the letter $i_1$ the deflation of which yielded $i_0$.
This $i_1$, now as a letter of ${\cal T}_{\gamma_{\epsilon}-a_{\epsilon}}$,
shall be the letter
to start with the next step leading to an edge
$f_2^{(l_2)}$ with $s(f_2^{(l_2)})=r(f_1^{(l_1)})$.
$f_1^{(l_1)}, f_2^{(l_2)}$ defines a path of length $2$ on
$\Sigma_{\epsilon}$.
The infinite path assigned to $i_0$ is obtained by infinite repetition
of that step.

There are exactly two paths obtained by this procedure which are constant,
namely the one assigned
to the letter to the right as well as the one to the left of $\vec{0}$.
We denote them by $\xi_+$ and $\xi_-$ resp..
Any other path obtained from a letter
of ${\cal T}$ will up to finitely many elements agree with
$\xi_+$ or $\xi_-$ depending on whether it has been chosen
right or left of $\vec{0}$. Moreover,
different letters are mapped onto different paths, and any path
ultimately agreeing with $\xi_+$ or $\xi_-$ can be obtained from
a letter of ${\cal T}$ (a preimage is easily found by inversion of the
above process). To summarize, since the letters of $\cal T$
are in one to one correspondence of the elements of
the orbit through ${\cal T}$, the above procedure defines a
mapping $\phi$ from that orbit to the subspace of paths which
ultimately agree with $\xi_+$ or $\xi_-$.
To make contact with the results of \cite{HPS} the following
questions have to be clarified (the natural topology of
${\cal P}(\Sigma_{\epsilon})$, turning it into a totally disconnected
compact space, is given below).
\begin{enumerate}
\item
Is it possible to extend this mapping to the closure of that orbit, i.e.\
to all representatives of $\Psi_{\epsilon}$?
\item
Let ${\cal R}(\Sigma_{\epsilon}) = \phi(R_{\Omega_{\epsilon}})$.
Does $\phi$ induce a homeomorphism between the quotients
$\Omega_{\epsilon}/R_{\Omega_{\epsilon}}$ and
${\cal P}(\Sigma_{\epsilon})/{\cal R}(\Sigma_{\epsilon})$?
\end{enumerate}
These questions remain open in this work, but let us
comment on them:\\
1)
The first problem may possibly be attacked as follows:
For
${\cal T}'\in \Omega_{\epsilon}$ consider an increasing net of finite parts
$\eta_1\subset \eta_2 \subset\cdots\subset {\cal T}'$ representing words
which all contain the letter right of $\vec{0}$, such that
${\cal T}'$ is approximated by these parts, i.e.\
\begin{equation}
{\cal T}'=\bigcup_{n=1}^{\infty}\eta_n .
\end{equation}
As ${\cal T}'$ is locally isomorphic to ${\cal T}$, for each
$n$ there is a $\lambda_n\in\Lambda_{\epsilon}$, such that
\begin{equation}                               \label{seq5}
\eta_n\subset {\cal T}-\pi(\lambda_n),
\end{equation}
and we assign to the pair $(\eta_n,\lambda_n)$ the path which is given
by the letter to the right of $\pi(\lambda_n)$ in ${\cal T}$.
Although that specific letter will be the same for each
$\lambda_n$ satisfying (\ref{seq5}),
the letter-deflation to which it belongs may depend on $\lambda_n$.
If for each $n$ there is a finite number $b_n$ such that
the first $n$ elements of the path determined
by $(\eta_n,\lambda_n)$ are independent of $\lambda_n$ as long as
$\eta_n$ contains at least $b_n$ letters to the right and to the left
of $\vec{0}$, then
the path assigned to ${\cal T}'$ may be taken to be the limit of
the above paths in the
topology a base of which is given by following the open-closed sets:
Each set
is defined by a path of finite length $\xi$ over $\Sigma_{\epsilon}$,
namely it contains
all (infinite) paths the beginning of which coincide with $\xi$.
This way $\phi$ becomes a continous mapping.
By construction of the path the $b_n$'s exist if $b_1$ does
($b_n$ has to grow exponentially with $n$). For substitution
sequences the existence of such a $b_1$ is guaranteed if they are
recognizable, for the definition of recognizability see \cite{Q}.
For $A_4$, $D_5$, $A_7$ and $E_6$-projections
the existence of a $b_1$ follows from
$\partial(\pi^{\perp}(\gamma\cap \{E+a\}))\subset\pi^{\perp}(\Lambda)$.\\
2)
Provided the first problem is solved, surjectivity of $\phi$ follows from
the compactness of $\Omega_{\epsilon}$:
Any path $\{f^{l_i}_i\}_{i\geq 1}$ over $\Sigma_{\epsilon}$ may be
approximated by a sequence of paths $\xi_n$, where the first $n$ elements
of $\xi_n$ coincide with $\{f^{l_i}_i\}_{1\leq i\leq n}$ whereas the rest
coincides ultimately with $\xi_+$.
Let ${\cal T}-\pi(\lambda_n)$ be a sequence of tilings such that
$\phi({\cal T}-\pi(\lambda_n))=\xi_n$.
As $\Omega_{\epsilon}$ is compact, this sequence has a convergent
subsequence,
the limit of which is a preimage of $\{f^{l_i}_i\}_{i\geq 1}$.
\bigskip

\noindent
Despite the abovementioned unsolved questions it seems to us quite
plausible
that the ordered $K_0$-group of
$C(\Omega_{\epsilon}) \times_{\varphi_{\epsilon}} \Integer$
coincides - similar to the case in \cite{HPS} - with (\ref{k1},\ref{b10}),
the one of $AF(\Sigma_{\epsilon})$. As in \cite{HPS} we expect
${\cal R}(\Sigma_{\epsilon})$ to be generated by
${\cal R}_0(\Sigma_{\epsilon}) = \{(\xi,\xi')|\exists n\forall i\geq n:
\xi_i = \xi'_i\} $ together with the element $(\xi_+,\xi_-)$, so that
$AF(\Sigma_{\epsilon})\cong C^*({\cal R}_0(\Sigma_{\epsilon}))$ and
$C(\Omega_{\epsilon}) \times_{\varphi_{\epsilon}} \Integer\cong
C^*({\cal R}(\Sigma_{\epsilon}))$.
This is also supported by the observation that the ($1$-dimensional)
aperiodic tiling
contains already all elements needed for the construction of the
ordered $K_0$-group. In fact the $K_0$-group corresponds
to the lattice $\Lambda_{\epsilon}$ and the order relation may be expressed
with the help of the orthogonal projection onto $P_{\epsilon}$, the latter
being the subspace along which the strip is constructed.
\bigskip

\noindent
Our final remark concerns
the question after a connection between
the crossed product of the $2$-dimensional system (\ref{ps2})
and $AF(\Sigma_1\times\Sigma_2) \cong
AF(\Sigma_1)\otimes AF(\Sigma_2)$,
the path algebra over $\Sigma_1\times\Sigma_2$.
The latter $AF$-algebra is expected to be related to
the alternative description
of the $2$-dimensional quantum space (\ref{psi}) analoguously to the
$1$-dimensional case.
The directed system given by the Bratteli diagram of paths over
$\Sigma_1\times\Sigma_2$ is,
for $n\geq 0$
\begin{eqnarray}
 K_0(A_n) & = & \Lambda_1\otimes\Lambda_2 \cong \Integer^{r_1 r_2} ,\\
  h_{n\,*} & = & \sigma_1\otimes\sigma_2 ,
\end{eqnarray}
$K_0(A_{-1}) = \Integer$ and $h_{-1\,*} = (11\cdots 1)^T$,
an $r_1 r_2\times 1$-matrix, $r_{\epsilon} = dim(\Lambda_{\epsilon})$.
Hence
\begin{equation}                                 \label{ps3a}
K_0(AF(\Sigma_1\times\Sigma_2)) \cong \Lambda_1\otimes\Lambda_2 ,
\end{equation}
and
\begin{equation}                                 \label{ps4}
K_0^+(AF(\Sigma_1\times\Sigma_2)) =
\{x_1\otimes x_2\in \Lambda_1\otimes\Lambda_2|
(\vec{\nu}_1,x_1)(\vec{\nu}_2,x_2) > 0\}\cup\{0\} .
\end{equation}
Again $tr_*:K_0(AF(\Sigma_1\times\Sigma_2))\longrightarrow\Real$ is given
up to normalization by the scalar product of $x$
with the Perron Frobenius vector $\vec{\nu}_1\otimes\vec{\nu}_2$
of $\sigma_1\otimes\sigma_2$.
Hence, although the $K_0$-groups of the
the double crossed product (\ref{ps2}) related to the tiling
on the one hand and the above AF-algebra on the
other do not coincide, their ranges of
$tr_*$ would do so (compare with (\ref{bb1}) and the sequel), provided the
the above discussed equality
$K_0(C(\Omega_{\epsilon})\times_{\varphi_{\epsilon}}\Integer)
= K_0(AF(\Sigma_{\epsilon}))$ as ordered groups
holds.

Normalizing $\vec{\nu}_{\epsilon}$ by
$\vec{\nu}=\vec{p}(\tau)$ ($\tau = 2cos\frac{\pi}{h}$),
it follows from (\ref{pol3}) that for $A_{2n}$ (for which
$(\vec{\nu}_1,\Lambda_1)=(\vec{\nu}_2,\Lambda_2)$)
the elements of $(\vec{\nu}_1,\Lambda_1)$ form a subring of $\Real$, i.e.\
$(\vec{\nu}_1,z)(\vec{\nu}_1,z')\in (\vec{\nu}_1,\Lambda_1)$.
In case of $A_{2n+1}$ this holds only
for the even polynomials, namely only
$(\vec{\nu}_1,\Lambda_1)$ forms a subring of $\Real$.
This ring structure depends upon the normalization, and, if one has
in mind that the elements $(\vec{\nu}_{\epsilon},\alpha)$,
$\alpha\in {\cal L}_{\epsilon}$,
yield relative frequencies of prototiles,
one might prefer the normalization to be in a way that the sum of the
entries of
$\vec{\nu}_{\epsilon}$ equals $1$. For $A_{2n}$
this amounts to setting $\vec{\nu}=(2-\tau)\vec{p}(\tau)$
which leads to the same subring of $\Real$.
We conclude that, concerning the
values of $tr_*$, Connes' approach to the Penrose tilings and this approach
to 2-dimensional $A_4$-projections lead to the same result.
For $A_{2n+1}$ however the latter normalization yields
$\vec{\nu}_{2k}=\frac{4-\tau^2}{2\tau}\vec{p}_{2k}(\tau)$ and
$\vec{\nu}_{2k+1}=\frac{4-\tau^2}{4}\vec{p}_{2k+1}(\tau)$ both of
which do not generate $(\vec{\nu}_1,\Lambda_1)$ resp.\
$(\vec{\nu}_2,\Lambda_2)$ as a ring ($2$ is not invertible).
\bigskip

\noindent {\bf Acknowledgements}
\medskip

\noindent
I want to thank W. Nahm, A. Recknagel,
M. R\"osgen and M. Terhoeven for various fruitful discussions.
I would also like to express my gratitude to A. van Elst and J. Cuntz for
a conversation which encouraged me to compute the Pimsner-Voiculescu
exact sequence for the iterated crossed product.
Finally I want to thank J. Bellissard, A. Bovier and J.M. Ghez
for interesting
discussions, their hospitality during my stay at CPT-Marseille last
November, and for acquainting me with reference \cite{HPS}.

\noindent
{\bf Figure 1:}
The above figure is obtained by the strip method
as follows:
The incomplete cartwheel tiling corresponds to a
$\Integer^5$-projection
(onto $P_{\omega'}$ with ${\cal L}=\{\epsilon_i\}_{1\leq i\leq 5}$) taking
$a = \vec{0}$ (it is singular).
Two of the above parallels are obtained by projection out of the lattice
$<{\cal L}'> -\frac{1}{2}\rho$ onto the same plane $P_{\omega'}$,
where $<{\cal L}'>$ is the sublattice
of $\Integer^5$,
which is generated by
${\cal L}'=
\{{\cal C}^2\delta^{-1}(\epsilon_i-\epsilon_{i+1})\}_{1\leq i\leq 4}$,
${\cal C}$ is the connectivity matrix of $A_4$ and
$\rho = {\cal C}^2\delta^{-1}(\epsilon_5-\epsilon_1)$.
Again (the singular choice) $a = \vec{0}$ is taken.
Now the Ammann-quasycristal is up to its asymmetric part completed by
rotating around $\frac{2\pi}{5}$.
\bigskip

\noindent
{\bf Figure 2:}
Selfsimilarity of the
$\Integer^7$-projection.

\end{document}